\documentclass[12pt, journal, onecolumn, final, letter]{IEEEtran}

\usepackage{cite}
\usepackage{amsmath,amssymb,amsfonts}
\usepackage{algorithmic}
\usepackage{graphicx}
\usepackage{textcomp}
\usepackage{xcolor}
\def\BibTeX{{\rm B\kern-.05em{\sc i\kern-.025em b}\kern-.08em
    \kern-.1667em\lower.7ex\hbox{E}\kern-.125emX}}

\usepackage[utf8]{inputenc} 
\usepackage[T1]{fontenc}
\usepackage{url}
\usepackage{bm,bbm}
\usepackage{enumerate}
\DeclareMathOperator{\rank}{rank}
\DeclareMathOperator*{\error}{err}
\DeclareMathOperator*{\trace}{tr}
\DeclareMathOperator{\weight}{wt}

\usepackage{amsthm}
\theoremstyle{plain}
\newtheorem{theorem}{Theorem}
\newtheorem{lemma}{Lemma}

\newtheorem{definition}{Definition}
\newtheorem{remark}{Remark}

\usepackage{hyperref}
\allowdisplaybreaks

\begin{document}
\title{Soft BIBD and Product Gradient Codes} 

\author{
Animesh Sakorikar and Lele Wang \\
University of British Columbia\\
5500 Main Mall, Vancouver, BC V6T1Z4, Canada\\
Emails: \{animeshs, lelewang\}@ece.ubc.ca
}

\maketitle

\begin{abstract}
Gradient coding is a coding theoretic framework to provide robustness against slow or unresponsive machines, known as stragglers, in distributed machine learning applications. Recently, Kadhe et al. proposed a gradient code based on a combinatorial design, called balanced incomplete block design (BIBD), which is shown to outperform many existing gradient codes in worst-case adversarial straggling scenarios~\cite{Kadhe--Koyluoglu--Ramchandran2019}. However, parameters for which such BIBD constructions exist are very limited~\cite{Colbourn--Dinitz2006}. In this paper, we aim to overcome such limitations and  construct gradient codes which exist for a wide range of system parameters while retaining the superior performance of BIBD gradient codes. Two such constructions are proposed, one based on a probabilistic construction that relax the stringent BIBD gradient code constraints, and the other based on taking the Kronecker product of existing gradient codes. The proposed gradient codes allow flexible choices of system parameters while retaining comparable error performance.
\end{abstract}

\section{Introduction}
Due to recent increases in size of available training data, a variety of machine learning tasks are distributed across multiple computing nodes to speed up the learning process. 
However, the theoretical speedup from distributing computations and parallelization may not be achieved in practise due to slow or unresponsive computing nodes, known as \emph{stragglers}. Stragglers can significantly impact the accuracy and efficiency of the computation~\cite{Chen2017, Tandon2017, Yadwadkar--Hariharan--Gonzalez2016}. Therefore, mitigating stragglers is one of the major challenges in the design of distributed machine learning systems. 

Basic approaches to mitigate stragglers include ignoring stragglers, detecting and avoiding stragglers~\cite{Yadwadkar--Hariharan--Gonzalez2016}, or replicating tasks across computing nodes~\cite{Chen2017},~\cite{Wang--Joshi--Wornell2015}. Recently, several coding theoretic based schemes were proposed to allow  more efficient use of redundant computations in mitigating stragglers~\cite{Aktas--Peng--Soljanin2017, Lee--Lam--Pedarsani--Papailiopoulos--Ramchandran2018, Li--Mousavi--Avestimehr--Soltanolkotabi2018, Yu--Ali--Avestimehr2020}.

\emph{Gradient coding} is one of the coding theoretic schemes, proposed by~\cite{Tandon2017}, to mitigate stragglers. In a gradient coding scheme, a central processor partitions the uncoded training data into pieces and distributes them among multiple computing nodes, called workers. Each worker computes a linear combination of the gradients on its assigned training data pieces. The central processor aggregates the returned linear combinations to obtain the gradient sum. The goal is to design the set of linear combinations such that the gradient sum can be recovered \emph{exactly} when any $s$ set of linear combinations are lost. Several gradient codes under the \emph{exact recovery} criterion are proposed in~\cite{Tandon2017, Halbawi--Azizan--Salehi--Fariborz2018, Reisizadeh--Prakash--Pedarsani--Avestimehr2019}. 

While exact recovery is a natural criterion that extends from traditional coding theory, it has some limitations for distributed learning applications. Firstly, the computational loads for exact recovery are proportional to the number of stragglers~\cite{Tandon2017}, and therefore an accurate estimate of the number of stragglers is needed to minimize unnecessary computation. Moreover, in many learning algorithms, such as stochastic gradient decent, an approximation of the actual gradient sum is sufficient~\cite{Raviv--Tandon--Dimakis--Tamo2018}. For these reasons, it is common to consider gradient codes that \emph{approximately} recover the gradient sum such that the squared $2$-norm of the difference between the actual gradient sum and its approximation is minimized, referred to as \emph{squared error}~\cite{Raviv--Tandon--Dimakis--Tamo2018}. A variety of approximate gradient code constructions are proposed in~\cite{Charles--Papailiopoulos--Ellenberg2017, Raviv--Tandon--Dimakis--Tamo2018, Charles--Papailiopoulos2018, Kadhe--Koyluoglu--Ramchandran2019, Wang--Charles--Papailiopoulos2019, Wang--Liu--Shroff2019, Glasgow--Wootters2020, Aktas--Peng--Soljanin2017, Sarmasarkar--Lalitha--Karamchandani2021}. 

In approximate gradient coding, common assumptions on the straggling scenarios include (i) random straggling, where stragglers are assumed to follow some stochastic assumptions, and (ii) worst-case straggling with no stochastic assumptions, where any $s$ subset of workers can be straggled. 
In this paper, we consider the squared error in the worst-case straggling scenario, referred to as \emph{worst-case squared error}, because it may be difficult to estimate the statistical model of stragglers in real time for some applications, such as massive-scale elastic and serverless systems~\cite{Kadhe--Koyluoglu--Ramchandran2019}. 

Under this performance criterion, we discuss two existing gradient codes: fractional repetition codes (FRCs)~\cite{Tandon2017} and balanced incomplete block design (BIBD) gradient codes~\cite{Kadhe--Koyluoglu--Ramchandran2019}. An FRC is constructed by partitioning workers and gradients into equally sized subsets, such that each worker in a partition is assigned every gradient in the same gradient partition. FRCs are easy to construct and exist for a wide range of distributed system parameters, but perform poorly in worst-case straggling scenarios~\cite{Charles--Papailiopoulos--Ellenberg2017}. BIBD gradient codes are constructed from combinatorial block designs, and are robust in worst-case straggling scenarios. However, the set of system parameters for which a BIBD gradient code is known to exist is very limited~\cite{Colbourn--Dinitz2006}. 

This motivates us to ask the following question: \emph{Is it possible to construct gradient codes that exist for a wide range of system parameters while retaining the superior worst-case performance produced by BIBD gradient codes?} We answer this affirmatively by providing two new constructions. 
In the first construction (Section~\ref{sec:PBIBD}), we propose a probabilistic gradient code construction that relaxes the stringent BIBD gradient code constraints, referred to as \emph{Soft BIBD} gradient codes. Unlike the BIBD gradient codes, which require constant computation load on each worker and constant number of shared computations between any pair of workers, in our construction, we only require these constraints to be satisfied \emph{on average}. 
As shown in Fig~\ref{fig:cmp_soft_comb_bibd} in Section~\ref{Probabilistic BIBD Gradient Code Generation}, our probabilistic construction enlarges the set of system parameters for which we can construct gradient codes.
Moreover, we show that the expected squared error of our construction is lower than that of a BIBD gradient code with the same system parameters if such a BIBD gradient code exists. 
In the second construction (Section~\ref{sec:KroneckerProduct}), we propose new gradient codes by taking the Kronecker product of existing gradient codes. Bounds on the normalized worst-case squared error are derived for four types of constructions: the Kronecker products of (i) an FRC with an FRC, (ii) an FRC with a BIBD, (iii) a BIBD with a BIBD, and (iv) a Soft BIBD with a Soft BIBD. 
The derived bounds guarantee that the Kronecker products of two BIBDs has comparable performance to the two component BIBDs. 

\section{Problem Formulation}
We use  bold capital script $\bm G$ for random matrices, capital script $G$ for deterministic matrices, and bold script $\bm f$ for vectors. Let $ I_{n}$ be the $n\times n$ identity matrix, $J_{k\times n}$ be the $k \times n$ all-one matrix, $\bm 1_k$ denote the $k\times 1$ all-one column vector, and $\bm 0_k$ denote the $k\times 1$ all-zero column vector. 

\subsection{Distributed Learning}
In many machine learning algorithms, the goal is to find a model $\bm \theta_\mathrm{min}$ that minimizes some loss function $L$ over a training data set. Specifically, given training data $\{(x_i,y_i)\}_{i=1}^k$, the aim is to compute
\begin{equation}
\bm \theta_\mathrm{min} = \arg\min_{\bm \theta \in \mathbbm{R}^p} \frac{1}{k}\sum_{i=1}^k L(\bm \theta, x_i,y_i). 
\end{equation}
Gradient descent is an iterative algorithm that estimates the optimal model $\bm \theta_\mathrm{min}$. Starting with some initial guess $\bm \theta^{(0)}$, the model $\bm \theta^{(t)}$ at iteration $t$ is updated as
\begin{equation}
\label{eqn:gradient_descent_alg}
\bm \theta^{(t+1)} = \bm \theta^{(t)} + \frac{\alpha}{k}\sum_{i=1}^k \nabla L(\bm \theta^{(t)}, x_i,y_i),
\end{equation}
where $\nabla L(\bm \theta^{(t)}, x_i,y_i)$ is the gradient of loss function $L$, and $\alpha$ is the learning rate. 

When the training data size $k$ is large, the gradient sum computation in \eqref{eqn:gradient_descent_alg} is a computational bottleneck. To avoid this computational bottleneck, we can distribute the $k$ gradient computations over multiple workers, known as distributed learning, or distributed gradient descent. In each iteration of distributed gradient descent, a central processor transmits the current model to each worker in the distributed system. Each worker computes some subset of the $k$ gradients, and returns the sum to the central processor. The central processor computes the gradient sum from the returned results, and updates the model as done in \eqref{eqn:gradient_descent_alg}. Henceforth we will focus on a single iteration of distributed gradient descent. 

\subsection{Gradient Coding}
\label{sec:model-GC}
Consider a distributed learning (or distributed gradient descent) setting with $n$ workers $\mathcal{W}$ and $k$ training data pieces $\{(x_i,y_i)\}_{i=1}^k$. 
The goal of gradient coding is to introduce redundancy in gradient computations so that the distributed system is robust against stragglers~\cite{Tandon2017}. 

A gradient code (GC) can be characterized by a $k\times n$ binary encoding matrix $G$. Row $i\in[k]:=\{1,2,\dotsc,k\}$ of $G$ corresponds to data piece $(x_i,y_i)$, and column $j\in[n]$ corresponds to worker $j$, where $G_{ij}=1$ if worker $j$ computes the gradient of $(x_i,y_i)$, and $G_{ij}=0$ otherwise. 
The worker load is the number of gradient computations assigned to a worker, and data redundancy is the number of workers computing the gradient of a data piece. When each worker has the same worker load $l$, and each data piece has redundancy $r$, $G$ is called an $(n,k,l,r)$-GC. 

The goal is to compute the gradient sum in~\eqref{eqn:gradient_descent_alg}. Since we consider only one iteration of distributed gradient descent, we omit the dependence on time $t$. Then the goal is to compute
\begin{equation*}
\sum_{i=1}^k \nabla L(x_i,y_i) = \bm f \bm 1_k,
\end{equation*}
where $\bm f = (\nabla L(x_1,y_1), \dotsc, \nabla L(x_k,y_k))$. 
Let $s$ be the number of straggling workers, $\mathcal{U} \subset \mathcal{W}$ be the set of non-straggling workers, and $G_\mathcal{U}$ be the sub-matrix of $G$ with columns indexed by $\mathcal{U}$. The central processor receives 
$
    \bm f G_\mathcal{U}
$
from the workers in $\mathcal{U}$. The goal is to recover $ \bm f \bm 1_k$ from $\bm f G_\mathcal{U}$ for any $\mathcal{U}$ such that $|\mathcal{U}|=n-s$. Notice that $\bm f G_\mathcal{U}$ is a $1\times(n-s)$ vector. Each entry of $\bm f G_\mathcal{U}$ is a result (partial gradient sum) returned by a non-straggling worker in $\mathcal{U}$. To approximate the gradient sum, we take a linear combination of returned results
\begin{equation*}
    \bm f G_\mathcal{U} \bm v, 
\end{equation*}
where $\bm v \in \mathbbm{R}^{n-s}$ is called the decoding vector.
Then the difference between the gradient sum approximation $\bm f G_\mathcal{U} \bm v$ and the target computation $\bm f \bm 1_k$ is 
\begin{equation}
\label{eqn:target_approx_diff}
    \bm f G_\mathcal{U} \bm v - \bm f \bm 1_k = \bm f (G_\mathcal{U} \bm v - \bm 1_k).
\end{equation}

Note that \eqref{eqn:target_approx_diff} depends on $\bm f$, which we do not know a priori. Thus we focus on minimizing the squared 2-norm of $(G_\mathcal{U}\bm v~-~\bm 1_k)$. We can now define the optimal decoding vector  as 
\begin{equation*}
\bm v_\mathrm{opt} (G,\mathcal{U}) := \arg \min_{\bm v \in \mathbbm{R}^{n-s}} \Vert G_\mathcal{U} \bm v - \bm 1_k \Vert_2^2.
\end{equation*}
Then the normalized worst-case squared error when $s$ workers are straggled is defined as
\begin{equation}
\label{eqn:error_defn}
\error(G,s) = \frac{1}{k} \max_{\substack{\mathcal{U}\subset\mathcal{W}\\ |\mathcal{U}|=n-s}} \Vert G_\mathcal{U} \bm v_\mathrm{opt} (G,\mathcal{U}) - \bm 1_k \Vert_2^2.
\end{equation}
This normalization allows us to compare between gradient codes with different number of gradient computations. 
Note that to make an objective comparison, one should choose gradient codes with the same fractional redundancy $r/n$, since each data piece is redundantly assigned to the same fraction of total workers in each gradient code being compared. We then compare the normalized worst-case squared errors as a function of the fraction of straggling workers. Note that $r/n = l/k$. Throughout this paper, we refer to the fractional redundancy as the density of the gradient code.

\subsection{Existing constructions}
In~\cite{Tandon2017}, the authors show that an $(n,k,l,r)$-GC can exactly recover all $k$ gradients for any set of $s$ stragglers if 
\begin{equation}
\label{eqn:exact_recovery}
l \geq \frac{k(s+1)}{n}.
\end{equation}
In the following, we focus on two constructions relating to the technical sections of our paper. 
In \cite{Tandon2017},  a gradient code construction called FRC is provided. An FRC with $n$ workers, $k$ data pieces, worker load $l$, and redundancy $r$ is called an $(n,k,l,r)$-FRC. The encoding matrix $G^\mathrm{F}$ is given by
\begin{equation*}
G^\mathrm{F} = 
\begin{bmatrix}
 J_{l\times r} &  0_{l\times r} & \dotsc &  0_{l\times r} \\
 0_{l\times r} &  J_{l\times r}  & \dotsc &  0_{l\times r} \\
\vdots & \vdots& \ddots & \vdots \\
 0_{l\times r} &  0_{l\times r} & \dotsc & J_{l\times r}
\end{bmatrix},
\end{equation*}
where $ 0_{l\times r} $ is the $l\times r$ all-zero matrix. An $(n,k,l,r)$-FRC can exactly recover all gradients if condition \eqref{eqn:exact_recovery} is satisfied. 
FRCs can also be used to approximately recover the gradient sum. The normalized worst-case squared error of an $(n,k,l,r)$-FRC $G^\mathrm{F} $ with $s$ stragglers is~\cite[Section 4.1]{Charles--Papailiopoulos--Ellenberg2017} 
\begin{equation}
\label{eqn:frc-err}
    \error(G^\mathrm{F},s) = \frac{l}{k} \left\lfloor \frac{s}{r} \right\rfloor.
\end{equation}
For completeness, we include a proof of~\eqref{eqn:frc-err} in Appendix~\ref{appendix:error_frc}.

Gradient codes can also be constructed from combinatorial balanced incomplete block designs (BIBD) as proposed in \cite{Kadhe--Koyluoglu--Ramchandran2019}. A combinatorial design is a pair $(\mathcal{X}, \mathcal{A})$, where $\mathcal{X} = \{x_1,x_2,\dotsc, x_k\}$ is a set of elements called points, and $\mathcal{A} = \{A_1, A_2,\dotsc, A_n\}$ is a collection of subsets of $\mathcal{X}$, called blocks. A design $(\mathcal{X}, \mathcal{A})$ is called a $(n,k,l,r,\lambda)$-BIBD if there are $k$ points in $\mathcal{X}$, $n$ blocks in $\mathcal{A}$, each of size $l$, where every point is contained in $r$ blocks, and any pair of distinct points is contained in exactly $\lambda$ blocks. We can represent a BIBD by an $k\times n$ incidence matrix $M$, where $M_{i,j} = 1$ iff $x_i\in A_j$ and $0$ otherwise. A BIBD is called \emph{symmetric} if its incidence matrix satisfies $M=M^T$. For any $(n,k,l,r,\lambda)$-BIBD with incidence matrix $M$, its dual-design is defined as the design with incidence matrix $M^T$.

A key observation in~\cite{Kadhe--Koyluoglu--Ramchandran2019} is that one can design a gradient code from a BIBD (with incidence matrix $M$) by setting the encoding matrix of the gradient code $G^\mathrm{B} = M$. Clearly the gradient code $G^\mathrm{B}$ is an $(n,k,l,r)$ gradient code. Moreover, if $G^\mathrm{B}$ is designed from a symmetric BIBD or the dual-design of a BIBD, then every pair of distinct workers of $G^\mathrm{B}$ share exactly $\lambda$ gradients to compute. This introduces redundancy and provides robustness against stragglers. Throughout this paper, we consider BIBD gradient codes, which are constructed from symmetric BIBDs and dual-designs of BIBDs.  We call these codes $(n,k,l,r,\lambda)$-BIBDs, or $(n,k,l,\lambda)$-BIBDs when the parameter $r$ is not relevant.
For convenience, the number of shared gradient computations between any pair of workers $\lambda$ is known as the number of intersections. 

BIBD gradient codes provide superior performance against worst-case stragglers: the optimal decoding vector and normalized worst-case squared error of a BIBD depend only on the number of stragglers, and not on the specific straggling scenario. Specifically, an $(n,k,l,\lambda)$-BIBD $G^\mathrm{B}$ with $s$ stragglers has a constant optimal decoding vector, and the normalized worst-case squared error of $G^\mathrm{B}$ with $s$ stragglers is given by~\cite[Theorem 1]{Kadhe--Koyluoglu--Ramchandran2019}
\begin{equation}
\label{eqn:bibd-error}
    \error(G^\mathrm{B},s) = 1 - \frac{l^2(n-s)}{kl+k\lambda(n-s-1)}.
\end{equation} 

For the same set of parameters $(n,k,l,r)$, an $(n,k,l,r,\lambda)$-BIBD promises better worst-case straggling performance than an $(n,k,l,r)$-FRC when it exists. However, the set of parameters $(n,k,l,r,\lambda)$ for which a BIBD is known to exist is very limited~\cite{Colbourn--Dinitz2006}. This motivates us to propose a new gradient code that exists for a wide range of parameters, while retaining the superior performance of BIBD gradient codes.

\section{Soft BIBD Gradient Codes}
\label{sec:PBIBD}
In this section, we propose a new gradient code, referred to as \emph{Soft BIBD} gradient codes. We generate the random encoding matrix $\bm G^\mathrm{PB}$  such that \emph{on average} the desired BIBD properties (such as worker load $l$ and $\lambda$ shared computations between any two workers) are satisfied. 
The construction of Soft BIBD  gradient codes is provided in Section~\ref{Probabilistic BIBD Gradient Code Generation}, which demonstrates that Soft BIBD gradient codes exist for a wider range of system parameters than combinatorial BIBD gradient codes.
Moreover in Section~\ref{Probabilistic BIBD Error} we show that Soft BIBD gradient codes have smaller average squared error than that of combinatorial BIBDs.

\subsection{Existence and Construction}
\label{Probabilistic BIBD Gradient Code Generation}
We define a joint distribution $p(x^n)$ on the $n$ entries of the first row in the encoding matrix $\bm G^\mathrm{PB}$ and generate all $k$ rows in the encoding matrix i.i.d. according to $p(x^n)$. We discuss the constraints $p(x^n)$ should satisfy to ensure the desired BIBD properties. 
\begin{enumerate}
\item In order for each column $j\in[n]$ of $\bm G^\mathrm{PB}$ to have $l$ ones on average, the marginal distribution on the entry in column $j$, denoted by $X_j$, must satisfy
\begin{equation}
\label{eqn:L}
\mathbbm{P}(X_j = 1) = \sum_{\substack{x^n\in\{0,1\}^n \\ x_j = 1}} p(x^n) = \frac{l}{k}.
\end{equation}
\item In order for each pair of columns $1 \le i<j \le n$ to have $\lambda$ intersections on average, the distribution on the entries in columns $i$ and $j$ must satisfy
\begin{equation}
\label{eqn:lambda}
\mathbbm{P}(X_i = X_j = 1) = \sum_{\substack{x^n\in\{0,1\}^n \\ x_i = x_j = 1}} p(x^n) = \frac{\lambda}{k}.
\end{equation}
\item Since $p(x^n)$ is a probability distribution, it must satisfy
\begin{align}
\sum_{x^n\in\{0,1\}^n} p(x^n) &= 1, \label{eqn:ltp} \\
p(x^n) &\ge 0 \text{ for all } x^n \in \{0,1\}^n. \label{eqn:non-negative}
\end{align}
\end{enumerate}

Now the problem of finding $p(x^n)$ reduces to finding a non-negative solution to the linear system $A\bm p = \bm b$. We arrange the linear system so that~\eqref{eqn:ltp} corresponds to first row of $A$, \eqref{eqn:L} corresponds to the next $n$ rows, and~\eqref{eqn:lambda} corresponds to the remaining $\binom{n}{2}$ rows of $A$. Observe that $A$ is an $\left( 1+n+\binom{n}{2} \right) \times 2^n $ binary matrix, whose entries are the coefficients of $p(x^n)$ in equations~\eqref{eqn:L},~\eqref{eqn:lambda}, and~\eqref{eqn:ltp}. The entry in row $i \in [2^n]$ of $\bm p$ is $p(x^n)$, where $x^n$ is the binary expansion of $i-1$. Lastly, $\bm b$ is a column vector with $\left( 1+n+\binom{n}{2} \right)$ entries, where the first entry is $1$, the next $n$ entries are $l/k$, and the remaining entries are $\lambda/k$. The system for size $n=3$ is shown below:
\begin{align}
\label{sys:N=3}
    \begin{bmatrix}
    1 & 1 & 1 & 1 & 1 & 1 & 1 & 1 \\
    0 & 1 & 0 & 1 & 0 & 1 & 0 & 1 \\
    0 & 0 & 1 & 1 & 0 & 0 & 1 & 1 \\
    0 & 0 & 0 & 0 & 1 & 1 & 1 & 1 \\
    0 & 0 & 0 & 1 & 0 & 0 & 0 & 1 \\
    0 & 0 & 0 & 0 & 0 & 1 & 0 & 1 \\
    0 & 0 & 0 & 0 & 0 & 0 & 1 & 1 \\
    \end{bmatrix}
    \begin{bmatrix}
    p(000) \\
    p(001) \\
    p(010) \\
    p(011) \\
    p(100) \\
    p(101) \\
    p(110) \\
    p(111) \\
    \end{bmatrix}
    =
    \begin{bmatrix}
    1 \\
    l/k \\
    l/k \\
    l/k \\
    \lambda/k \\
    \lambda/k \\
    \lambda/k \\
    \end{bmatrix}.
\end{align}

The main theorem of this section provides the parameters $n,k,l$ and $\lambda$ for which a non-negative solution $p(x^n)$ to the system $A \bm p = \bm b$ exists. 

\begin{theorem}
\label{thm:pbibd_param}
For all $k \geq l\geq \lambda$, a solution exists for the linear system $A \bm p = \bm b$ defined by equations~\eqref{eqn:L}, \eqref{eqn:lambda}, and~\eqref{eqn:ltp}. Additionally a non-negative solution to $A\bm p = \bm b$ exists for parameters $n,k,l,\lambda$ in the region given by
\begin{equation}
\label{eqn:soft_bibd_cond_region}
    \{2\lambda \ge l, k \ge 3l-2\lambda\} \cup \bigcup_{\tau=2}^n \left( \left \{k \geq n\left( l - \frac{n-1}{\tau}\lambda \right) \right \} \cap \left\{l \geq \lambda\left(\frac{n-1}{\tau - 1} \right) \right\} \right).
\end{equation}
\end{theorem}

\begin{figure}[htbp]
  \centering
  \includegraphics[width=0.9\linewidth]{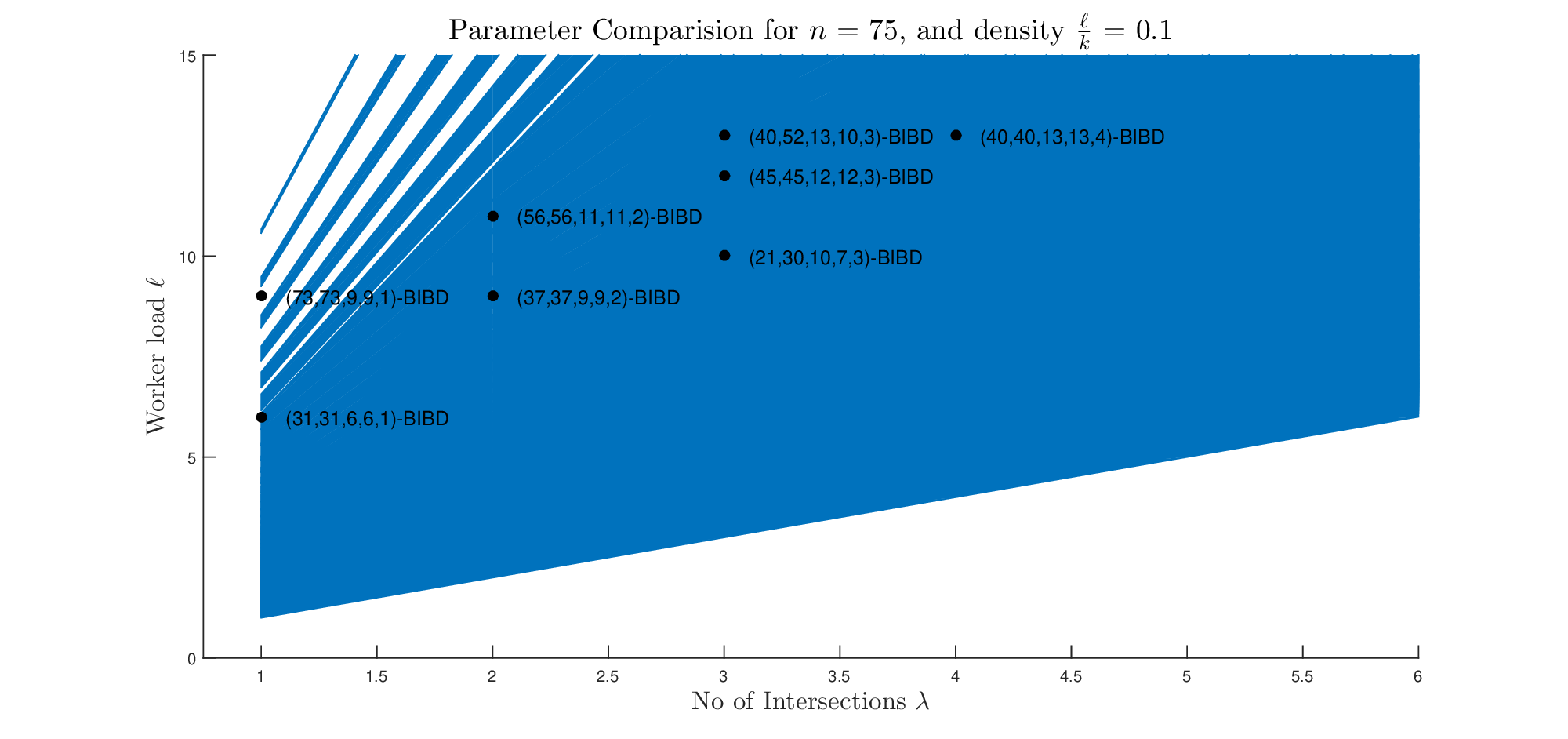}
  \caption{The region in blue is the valid parameter region for our proposed probabilistic gradient code construction. The black dots correspond to known combinatorial BIBD gradient codes with a similar density.}
  \label{fig:cmp_soft_comb_bibd}
\end{figure}

Theorem~\ref{thm:pbibd_param} shows that Soft BIBD gradient codes exist for a wider range of system parameters than combinatorial BIBD gradient codes. In Fig~\ref{fig:cmp_soft_comb_bibd}, we fix the number of workers $n = 75$ and the density $\tfrac{l}{k} = 0.1$. The region in blue is the parameter region for which a Soft BIBD can be generated as given by Theorem~\ref{thm:pbibd_param}. We also plot various combinatorial BIBDs with similar densities. As shown in Fig~\ref{fig:cmp_soft_comb_bibd}, Theorem~\ref{thm:pbibd_param} allows us to construct Soft BIBDs for a wide range of system parameters. Moreover, Soft BIBDs also exist for non-integer values for average worker load and average number of intersections. 

A key observation in establishing Theorem~\ref{thm:pbibd_param} is that matrix $A$ turns out to be the same as the generator matrix of a Reed-Muller (RM) code of order 2 and block length $2^n$, which are defined as follows \cite[p.~373]{MacWilliams--Sloane1977}. 

\begin{definition}
The binary Reed-Muller code of order $2$ and block length $2^n$ is the set of all vectors $f(v_1,\dotsc,v_n)$, where $f$ is a Boolean function which is a polynomial of degree at most $2$, and $(v_1,\dotsc,v_n)$ is any length $n$ binary vector. 
\end{definition} 

One can verify that the matrix in equation~\eqref{sys:N=3} constitutes a basis for the Reed-Muller code of order $2$ and length $2^3$, and therefore is a generator matrix for this code. In general, this equivalence is true for any $n$.

\begin{lemma} 
\label{lemma:rm_eq}
The matrix $A$ defined by constraints~\eqref{eqn:L},~\eqref{eqn:lambda}, and~\eqref{eqn:ltp} is a generator matrix of the Reed-Muller code of order $2$ and length $2^n$.
\end{lemma} 

Farkas' lemma is also key in establishing this result. We first state the lemma. 

\begin{lemma}[{Farkas' lemma~\cite[Section 5.8]{Boyd-Vendenberghe2004}}]
\label{lemma:farkas}
Let $A \in \mathbb{R}^{m \times n}$ and $\bm b \in \mathbb{R}^m$. Then exactly one of the following two assertions is true:
\begin{enumerate}
    \item There exists $\bm p \in \mathbb{R}^n$ such that $A \bm p = \bm b$ and $\bm p \ge \bm 0$ elementwise. 
    \item There exists $\bm y \in \mathbb{R}^m$ such that $A^T \bm y \geq \bm 0$ elementwise and $\bm b^T y < 0$.
\end{enumerate}
\end{lemma}

\begin{IEEEproof} [\bf Proof of Theorem~\ref{thm:pbibd_param}]
A solution to the system of equations $A \bm p = \bm b$ exists if and only if the rank of the augmented coefficient matrix $\Hat{A} = [A|\bm b]$ is equal to the rank of $A$. It is known that generator matrices for Reed-Muller codes have full rank, thus from Lemma~\ref{lemma:rm_eq} we have $\rank( A)=1+n+\binom{n}{2}$, and $\rank(\Hat{A}) \geq \rank(A)=1+n+\binom{n}{2}$. Since $\Hat{A}$ is a $\left (1+n+\binom{n}{2} \right ) \times \left (2^n + 1\right )$ matrix, $\rank (\Hat{A}) \leq 1+n+\binom{n}{2}$. Thus $\rank(\Hat{A})=\rank(A)$, and a solution to the system exists for all $n\geq l\geq\lambda$.

To show a non-negative solution exists when $2\lambda >l$ and $k \ge 3l - 2\lambda$, consider a solution of the form $\bm p = (\alpha, \beta, \dotsc, \beta, \gamma)$, where $p(0^n) = \alpha$, $p(1^n) = \gamma$, and $p(x^n) = \beta$ for all $x^n \not\in \{0^n, 1^n\}$. Then equations \eqref{eqn:ltp}, \eqref{eqn:L} and \eqref{eqn:lambda} become:  
\begin{align} 
\label{eqn:ltp_red}
    \alpha + \beta (2^n-2) + \gamma &= 1, \\
\label{eqn:L_red}
    \beta (2^{n-1} - 1) + \gamma &= \frac{l}{k}, \\
\label{eqn:lambda_red}
    \beta (2^{n-2} - 1) + \gamma &= \frac{\lambda}{k}.
\end{align} 
Solving equations \eqref{eqn:ltp_red}, \eqref{eqn:L_red}, and \eqref{eqn:lambda_red} for $\alpha,\beta$ and $\gamma$, we get
\begin{align*}
    \beta &= \frac{l-\lambda}{k(2^{n-2})}, \\
    \gamma &= \frac{1}{k}\left[2\lambda - l + \frac{l-\lambda}{2^{n-2}}\right], \\
    \alpha &= 1 + \frac{2\lambda - 3l}{k} + \frac{l-\lambda}{k (2^{n-2})}.
\end{align*}
For a positive solution $\bm p$ to the system, we require $\alpha,\beta$ and $\gamma$ to be non-negative. One can check they are indeed non-negative when $2\lambda\geq l$ and $k\geq 3l-2\lambda$.

We now show that a non-negative solution also exists in the region given by
\begin{equation}
\label{eqn:farkas_regions}
    \bigcup_{\tau=2}^n \left(\left\{k \geq n\left( l - \frac{n-1}{\tau}\lambda \right)\right\} \bigcap \left\{l \geq \lambda\left(\frac{n-1}{\tau - 1} \right) \right\} \right).
\end{equation}
To this end, we restrict the solution space and consider a solution with a specific structure. Let $S_\tau = \{x^n: \weight(x^n) = \tau \}$, where $\weight(x^n)$ is the hamming-weight of the binary sequence $x^n$. We require that for any $\tau \in [n]\cup\{0\}$ and for all $x^n \in S_\tau$, we have 
\begin{equation*}
    p(x^n) = \alpha_\tau, \quad 0\leq \alpha_\tau\le 1.
\end{equation*} 
Observe that for all $\tau \in [n]\cup\{0\}$, each $x^n \in S_\tau$ appears exactly once in~\eqref{eqn:ltp}, and $|S_\tau| = \binom{n}{\tau}$. Thus under the above restriction, equation~\eqref{eqn:ltp} simplifies to
\begin{equation}
\label{eqn:weight_class_ltp}
    \alpha_0 + n\alpha_1 + \sum_{\tau=2}^n \binom{n}{\tau}\alpha_\tau = 1
\end{equation}
Similarly, by symmetry, there are exactly $\binom{n-1}{\tau-1}$ elements of $S_\tau$ in any equation in the form of equation~\eqref{eqn:L}, and $\binom{n-2}{\tau-2}$ elements of $S_\tau$ in any equation of the form given by equation~\eqref{eqn:lambda}. Therefore, under the above restriction, equations~\eqref{eqn:L} and~\eqref{eqn:lambda} simplify to
\begin{align}
    \label{eqn:weight_class_l}
    \alpha_1 + \sum_{\tau=2}^n \binom{n-1}{\tau-1}\alpha_\tau &= \frac{l}{k}, \\
    \label{eqn:weight_class_lambda}
    \sum_{\tau=2}^n \binom{n-2}{\tau-2}\alpha_\tau &= \frac{\lambda}{k}.
\end{align}
Thus by restricting the solution space, the problem of finding a non-negative solution to $A \bm p = \bm b$ reduces to solving the linear system of equations $W \bm \alpha = \bm z$ given by equations~\eqref{eqn:weight_class_ltp},~\eqref{eqn:weight_class_l}, and~\eqref{eqn:weight_class_lambda}. Note that $W$ is of the form
\begin{equation*}
    W =
    \begin{bmatrix}
    \binom{n}{0} & \binom{n}{1} & \binom{n}{2} & \cdots & \binom{n}{n-1} & \binom{n}{n} \\
    0 & \binom{n-1}{0} & \binom{n-1}{1} & \cdots & \binom{n-1}{n-2} & \binom{n-1}{n-1} \\
    0 & 0 & \binom{n-2}{0} & \cdots & \binom{n-2}{n-3} & \binom{n-2}{n-2} \\
    \end{bmatrix},
\end{equation*}
$\bm \alpha = (\alpha_0, \alpha_1, \alpha_2, \dotsc, \alpha_n )^T$, and $\bm z = (1, \tfrac{l}{k}, \tfrac{\lambda}{k})^T$.
We find conditions on $n,k,l$ and $\lambda$ such that the second statement of Farkas' lemma is not true for the system $W\bm\alpha = \bm z$. Then, under these conditions, the first statement of Farkas' lemma is true and there exists a non-negative solution to the system $W\bm\alpha = \bm z$. 
We find the aforementioned conditions region by contradiction. To this end, assume the second statement of Farkas' lemma  is true. Then there exists $\bm y = (y_1, y_2, y_3)$ satisfying $W^T \bm y \geq \bm 0$ elementwise and $\bm z^T y < 0$. Equivalently, there exists $\bm y = (y_1, y_2, y_3)$ satisfying 
\begin{align}
    \label{eqn:wt_class_cond1}
    y_1 &\ge 0, \\
    \label{eqn:wt_class_cond2}
    ny_1 + y_2 &\ge 0, \\
    \label{eqn:wt_class_cond3}
    \binom{n}{\tau}y_1 + \binom{n-1}{\tau-1}y_2 + \binom{n-2}{\tau-2}y_3 &\ge 0, \quad  \forall \tau\in\{2,\dotsc,n\}, \\
    \nonumber
    y_1 + \frac{l}{k}y_2 + \frac{\lambda}{k}y_3 &< 0.
\end{align}
Now aiming for a contradiction, for any $\tau\in\{2,\dotsc,n\}$, we have
\begin{align}
    \nonumber
    0 &> y_1 + \frac{l}{k}y_2 + \frac{\lambda}{k}y_3 \\
    \nonumber
    &\stackrel{(a)}{\ge} y_1 + \frac{l}{k}y_2 + \frac{\lambda}{k} \left( \frac{-\binom{n}{\tau}y_1 - \binom{n-1}{\tau-1}y_2}{\binom{n-2}{\tau-2}} \right) \\
    \nonumber
    &\stackrel{(b)}{\ge} y_1\left(1 - \frac{\lambda \binom{n}{\tau}}{k\binom{n-2}{\tau-2}} \right)  + (-n y_1) \left( \frac{l}{k}  - \frac{\lambda \binom{n-1}{\tau-1}}{k\binom{n-2}{\tau-2}}\right) \\
    \nonumber
    &= y_1 \left( 1 - \frac{nl}{k} + \frac{\lambda n(n-1)}{k\tau} \right) \\
    \label{eqn:farkas_contradiction}
    &:= y_1 \psi.
\end{align}
Above, $(a)$ follows from~\eqref{eqn:wt_class_cond3}. Moreover, $(b)$ follows from~\eqref{eqn:wt_class_cond2} and  is true when
\begin{equation}
\label{eqn:prob_region_parameter1}
    \frac{l}{k}  - \frac{\lambda \binom{n-1}{\tau-1}}{k\binom{n-2}{\tau-2}} > 0.
\end{equation}
From~\eqref{eqn:wt_class_cond1}, $y_1 \ge 0$. Therefore, if we have parameters $n,k,l$ and $\lambda$ such that $\psi \ge 0$, then $y_1\psi\ge 0$, which contradicts~\eqref{eqn:farkas_contradiction}.
Then by Farkas' Lemma, for these parameters $n,k,l$ and $\lambda$ there exists a non-negative solution to $W \bm \alpha  = \bm z$. Indeed, $\psi \ge 0$ if
and only if 
\begin{equation}
\label{eqn:prob_region_parameter2}
    k \ge n\left( l - \frac{\lambda(n-1)}{\tau} \right).
\end{equation}
Our analysis above holds if both ~\eqref{eqn:prob_region_parameter1} and~\eqref{eqn:prob_region_parameter2} are satisfied for any $\tau\in\{2,\dotsc,n\}$, which gives us the region in~\eqref{eqn:farkas_regions}.
\end{IEEEproof}

\subsection{Error of Probabilistic Gradient Codes}
\label{Probabilistic BIBD Error}
In this section, we analyze the normalized worst-case squared error performance of Soft BIBD gradient codes. Let $n,k,l$ and $\lambda$ satisfy the conditions in Theorem~\ref{thm:pbibd_param}, and $\bm G^\mathrm{PB}$ be an $n$ worker and $k$ gradient Soft BIBD gradient code, with  worker load $l$ and intersections $\lambda$ in expectation. Then $\bm G^\mathrm{PB}$ is called an $(n,k,l,\lambda)$ Soft BIBD . 
Since $\bm G^\mathrm{PB}$ is a randomly generated matrix, the performance depends on the specific realization. 
The following theorem characterizes the performance of an $(n,k,l,\lambda)$ Soft BIBD with $s$ stragglers.
\begin{theorem}
\label{theorem:pbibd_error}
Let $n,k,l$ and $\lambda$ lie in the region given by~\eqref{eqn:soft_bibd_cond_region}. For a probabilistic $(n,k,l,\lambda)$-BIBD gradient code $\bm G^\mathrm{PB}$ generated from the probability distribution specified in Section~\ref{Probabilistic BIBD Gradient Code Generation}, and for any set $\mathcal{U}$ of size $n-s$ non-stragglers, we have
\begin{equation}
\label{eqn:thm_pbibd_error}
    \frac{1}{k} \mathbbm{E} \left[ \Vert \bm G^\mathrm{PB}_\mathcal{U} \bm v_\mathrm{opt} (\bm G^\mathrm{PB},\mathcal{U}) - \bm 1_k \Vert_2^2 \right] \le 1 - \frac{l^2(n-s)}{kl+k\lambda(n-s-1)},
\end{equation}
where $s \in [n]$ is the number of stragglers.
\end{theorem}

\begin{figure}[htbp]
  \centering
  \includegraphics[width=0.75\linewidth]{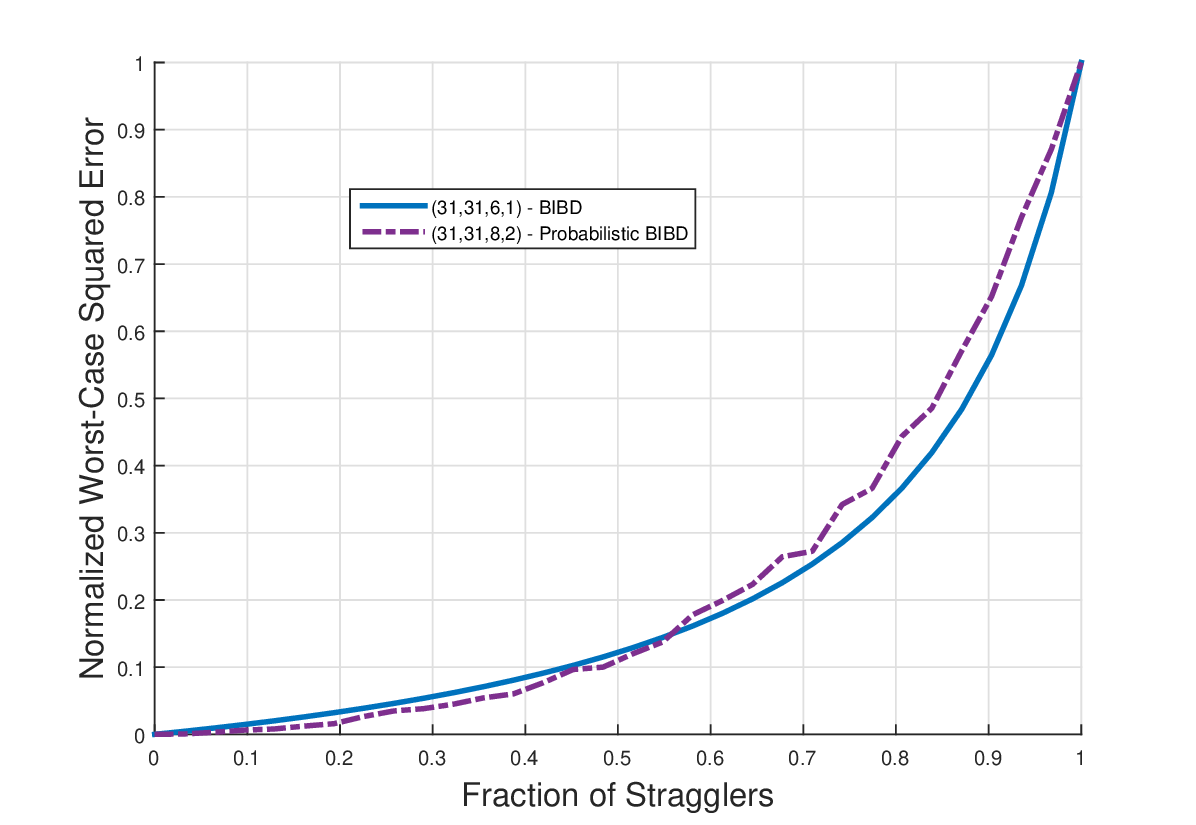}
  \caption{Comparison of Worst-case Normalized Errors of Combinatorial BIBDs and Soft BIBDs}
  \label{fig:comb_bibd_vs_soft_bibd_err}
\end{figure}

In the case that a combinatorial BIBD $G^\mathrm{B}$ with the same system parameters as $\bm G^\mathrm{PB}$ exists, the right hand side of~\eqref{eqn:thm_pbibd_error} is equal to $\error(G^\mathrm{B}, s)$. Then, since~\eqref{eqn:thm_pbibd_error} holds for any set of $n-s$ non stragglers, Theorem~\ref{theorem:pbibd_error} shows that 
\begin{equation*}
    \error(\bm G^\mathrm{PB}, s) \triangleq \max_{\substack{\mathcal{U}\subset\mathcal{W}\\ |\mathcal{U}|=n-s}} \frac{1}{k} \mathbbm{E} \left[ \Vert \bm G^\mathrm{PB}_\mathcal{U} \bm v_\mathrm{opt} (\bm G^\mathrm{PB},\mathcal{U}) - \bm 1_k \Vert_2^2 \right] \le \error(G^\mathrm{B}, s).
\end{equation*}
This suggests that there exists a realization of $\bm G^\mathrm{PB}$ with superior error performance to a combinatorial BIBD $G^\mathrm{B}$.
The simulation in Fig.~\ref{fig:comb_bibd_vs_soft_bibd_err} plots the errors of a combinatorial BIBD and a Soft BIBD with similar densities. The Soft BIBD was straggled at random for 2000 trials, and decoded with the optimal decoding vector obtained by taking the pseudoinverse. The simulation provides a Soft BIBD realization with similar density and comparable error performance to the combinatorial BIBD. 

To prove Theorem~\ref{theorem:pbibd_error}, we first state the following result from~\cite{Kadhe--Koyluoglu--Ramchandran2019}, which gives the optimal decoding vector and normalized worst-case squared error of a BIBD with $s$ stragglers. 
\begin{lemma} 
\label{lemma:bibd_error}
[Theorem 1 in \cite{Kadhe--Koyluoglu--Ramchandran2019}] For a non-random $(n,k,l,\lambda)$-BIBD gradient code $G^\mathrm{B}$ with $s$ stragglers, the optimal decoding vector is 
\begin{equation*}
    \bm v_{\mathrm{opt}} (G^\mathrm{B}, s) = \left( \frac{l}{l + \lambda(n-s-1) }\right) \bm 1_{n-s} ,
\end{equation*}
and the worst-case squared error is 
\begin{equation}
\label{eqn:bibd_error_unnormalized}
\max_{\substack{\mathcal{U} \subset [n] \\ |\mathcal{U}|=n-s}} \Vert G^\mathrm{B}_\mathcal{U}\bm v_{\mathrm{opt}} (G^\mathrm{B}, \mathcal{U}) - \bm 1_k \Vert_2^2 = k - \frac{l^2(n-s)}{l+\lambda(n-s-1)}.
\end{equation}
\end{lemma}
Then dividing both sides of the worst-case squared error in equation \eqref{eqn:bibd_error_unnormalized} by $k$, we obtain the BIBD error expression in~\eqref{eqn:bibd-error}. 

To prove Theorem~\ref{theorem:pbibd_error}, we bound the normalized squared error of a $(n,k,l,\lambda)$ Soft BIBD gradient code by decoding with the optimal decoding vector of an $(n,k,l,\lambda)$-BIBD provided in Lemma~\ref{lemma:bibd_error}.

\begin{IEEEproof}[\bf Proof of Theorem~\ref{theorem:pbibd_error}]
Since $\bm G^\mathrm{PB}$ is a random matrix with probability distribution specified in Section~\ref{Probabilistic BIBD Gradient Code Generation}, the expected number of ones in each column of  is $l$, and the expected number of intersections between any two columns is $\lambda$. Then for any set of $n-s$ non-straggling worker indices $\mathcal{U}$
\begin{align}
\label{eqn:pbibd_expectation_ones}
\mathbbm{E}[(\bm G^\mathrm{PB}_\mathcal{U})^T \bm 1_k] &= l \bm 1_{n-s} , \\
\label{eqn:pbibd_expectation_intersections}
\mathbbm{E}[(\bm G^\mathrm{PB}_\mathcal{U})^T (\bm G^\mathrm{PB}_\mathcal{U})] &= (l-\lambda)  I_{n-s} + \lambda  J_{(n-s)\times (n-s)}.
\end{align}
Define the constant decoding vector 
\[
\tilde{\bm v} =  \left( \frac{l}{l + \lambda(n-s-1) }\right) \bm 1_{n-s}.
\]
Then 

\begin{align*}
&\phantom{=} \frac{1}{k}  \mathbbm{E} [\Vert \bm G^\mathrm{PB}_\mathcal{U}\bm v_{\mathrm{opt}} (\bm G^\mathrm{PB}, \mathcal{U}) - \bm 1_k \Vert_2^2] \\
&= \frac{1}{k}  \left[ \sum_g \mathbbm{P}(\bm G^\mathrm{PB} = g) \Vert g_\mathcal{U}\bm v_{\mathrm{opt}} (g, \mathcal{U}) - \bm 1_k \Vert_2^2  \right] \\
&\stackrel{(a)}{\le} \frac{1}{k}  \left[ \sum_g \mathbbm{P}(\bm G^\mathrm{PB} = g) \Vert g_\mathcal{U}\tilde{\bm v} - \bm 1_k \Vert_2^2  \right] \\
&= \frac{1}{k}  \mathbbm{E}\left[\Vert \bm G^\mathrm{PB}_\mathcal{U} \tilde{\bm v} - \bm 1_k \Vert_2^2 \right] \\
&=\frac{1}{k}  \bigg(\mathbbm{E}\left[\bm 1_k^T \bm 1_k\right] - 2\tilde{\bm v}^T \mathbbm{E}\left[(\bm G^\mathrm{PB}_\mathcal{U})^T \bm 1_k \right] + \tilde{\bm v}^T \mathbbm{E}\left[(\bm G^\mathrm{PB}_\mathcal{U})^T \bm G^\mathrm{PB}_\mathcal{U} \right]\tilde{\bm v}  \bigg) \\[5 pt]
&\stackrel{(b)}{=}\frac{1}{k}  \left( k - 2l  \tilde{\bm v}^T \bm 1_{n-s} + \tilde{\bm v}^T ((l-\lambda) I_{n-s} + \lambda J_{n-s})\tilde{\bm v} \right) \\
&= 1 - \frac{l^2(n-s)}{kl+k\lambda(n-s-1)},
\end{align*}
where $(a)$ follows because $\tilde{\bm v}$ is a constant decoding vector that may not be optimal, and $(b)$ follows from equations \eqref{eqn:pbibd_expectation_ones} and \eqref{eqn:pbibd_expectation_intersections}.
\end{IEEEproof}

\section{Kronecker Product Gradient Codes}
\label{sec:KroneckerProduct}
In the previous section, we constructed probabilistic gradient codes that satisfy the desired BIBD properties on average. We now switch to a different construction, referred to as \emph{product gradient codes}, which are constructed by taking the Kronecker product of matrices of existing gradient codes. 
We first define the Kronecker product. 

\begin{definition}
If $A$ is a $k_1 \times n_1$ matrix, and $B$ is a $k_2\times n_2$ matrix, then the Kronecker product $A \otimes  B$ is a $k_1k_2 \times n_1n_2$ matrix given by 
\[
 A \otimes B :=
\begin{bmatrix}
A_{1,1} B & \cdots & A_{1,n_1}  B \\
\vdots & \ddots & \cdots \\
A_{k_1, 1} B & \cdots & A_{k_1,n_1} B \\
\end{bmatrix} .
\]
\end{definition}

We first state the following important lemma, which states that permuting the order in which the Kronecker product of gradient codes is taken does not affect the resulting error expression. 

\begin{lemma}
\label{lemma:permutation_equivalence}
Let $G^{(i)}$ be gradient codes with $n_i$ workers and $k_i$ gradients to compute for $i = 1,2$. Then for any $s \in [n_1n_2]$,
\begin{equation*}
    \error(G^{(1)} \otimes G^{(2)} , s) = \error(G^{(2)} \otimes G^{(1)}, s). 
\end{equation*}
\end{lemma}
The proof is deferred to Appendix~\ref{appendix:permutation_equivalence}.

We establish the normalized worst-case squared error performance for Kronecker products of two FRCs, an FRC with a BIBD, two BIBDs, and two Soft BIBDs in Theorems~\ref{thm:err_kron_frc}-\ref{thm:err_kron_prob_bibd}. 
For convenience, we extend the domain of FRC error expression in~\eqref{eqn:frc-err} to the set of real numbers. For an $(n,k,l,r)$-FRC $G^\mathrm{F}$ and any real number $0 \le s\le n$, let
\begin{equation*}
    \error(G^\mathrm{F},s) = \frac{l}{k}\left\lfloor\frac{s}{r}\right\rfloor. 
\end{equation*}

\begin{theorem}
\label{thm:err_kron_frc}
Let $G^{\mathrm{F}_i}$ be an $(n_i,k_i,l_i,r_i)$-FRC for $i=1,2$. Then for any $s\in[n_1n_2]$,
\begin{equation*}
\error(G^{\mathrm{F}_1}\otimes G^{\mathrm{F}_2},s) 
= \frac{l_1}{k_1}\error\left(G^{\mathrm{F}_2}, \frac{s}{r_1}\right) 
= \frac{l_2}{k_2}\error\left(G^{\mathrm{F}_1}, \frac{s}{r_2}\right). 
\end{equation*}
\end{theorem}

\begin{theorem}
\label{thm:err-frc-bibd}
Let $G^\mathrm{F}$ be an $(n_1, k_1, l_1, r_1)$-FRC and $G^\mathrm{B}$ be an $(n_2, k_2, l_2, r_2,\lambda_2)$-BIBD. Then the error of $G^\mathrm{F}\otimes G^\mathrm{B}$ with $s$ stragglers is given by
\begin{equation*}
\error(G^\mathrm{F}\otimes G^\mathrm{B},s)  = \error\left(G^\mathrm{F}, \frac{s}{n_2}\right) + \frac{l_1}{k_1}\error\left(G^\mathrm{B},b\right),
\end{equation*}
where $b = \left\lfloor\tfrac{1}{r_1}\left(s-\left\lfloor\tfrac{s}{r_1 n_2}\right\rfloor r_1n_2\right)\right\rfloor$. 
\end{theorem}

\begin{theorem}
\label{thm:err_kron_bibd}
Let $G^\mathrm{B_i}$ be $(n_i,k_i,l_i,r_i,\lambda_i)$-BIBDs for $i=1,2$. The error of $G^\mathrm{B_1} \otimes G^\mathrm{B_2}$ with any $s\in[n_1n_2]$ stragglers is upper bounded as
\begin{equation*}
\error(G^\mathrm{B_1} \otimes G^\mathrm{B_2}, s) 
\le 1 - \frac{(l_1l_2)^2 (n_1n_2 - s)}{k_1k_2 \left( d+\lambda_1 \lambda_2(n_1n_2 - s) \right)},
\end{equation*}
where 
\begin{equation*}
    d= (l_1-\lambda_1)(l_2-\lambda_2)+n_2(l_1\lambda_2-\lambda_1\lambda_2) + n_1(\lambda_1l_2-\lambda_1\lambda_2).
\end{equation*}
\end{theorem}

\begin{theorem}
\label{thm:err_kron_bibd_lb}
Let $G^\mathrm{B_i}$ be $(n_i,k_i,l_i,r_i,\lambda_i)$-BIBDs for $i=1,2$. The error of $G^\mathrm{B_1} \otimes G^\mathrm{B_2}$ with any $s\in[n_1n_2]$ stragglers is lower bounded as
\begin{align*}
    \error(G^\mathrm{B_1} \otimes G^\mathrm{B_2}, s) 
    &\ge 
    \max_{\mathcal{F}} \frac{1}{k_1k_2}\left(k_1k_2 - 2c_1c_2 + d_1d_2 \right),
\end{align*}
where 
\begin{equation*}
    \mathcal{F} = \{(s_1, s_2) : s_1 \in [n_1], s_2\in[n_2], (n_1n_2 - s) = (n_1 -s_1)(n_2 - s_2)\},
\end{equation*}
and
\begin{equation*}
    c_i = \frac{l_i^2(n_i - s_i)}{l_i + \lambda_i(n_i - s_i - 1)}, \quad d_i = \frac{l_i^2((l_i + \lambda_i)(n_i-s_i)+\lambda_i(n_i-s_i)^2)}{(l_i + \lambda_i(n_i - s_i - 1))^2},\quad i = 1,2.
\end{equation*}
\end{theorem}

\begin{theorem}
\label{thm:err_kron_prob_bibd}
Let $\bm G^\mathrm{PB_i}$ be  $(n_i,k_i,l_i,r_i,\lambda_i)$ Soft BIBDs for $i=1,2$. The error of $\bm G^\mathrm{PB_1} \otimes \bm G^\mathrm{PB_2}$ with any $s\in[n_1n_2]$ stragglers is upper bounded as
\begin{equation*}
\error(\bm G^\mathrm{PB_1} \otimes \bm G^\mathrm{PB_2}, s) 
\le 1 - \frac{(l_1l_2)^2 (n_1n_2 - s)}{k_1k_2 \left( d+\lambda_1 \lambda_2(n_1n_2 - s) \right)},
\end{equation*}
where 
\begin{equation*}
    d= (l_1-\lambda_1)(l_2-\lambda_2)+n_2(l_1\lambda_2-\lambda_1\lambda_2) + n_1(\lambda_1l_2-\lambda_1\lambda_2).
\end{equation*}
\end{theorem}

\begin{figure}[h!]
  \centering
  \includegraphics[width=0.8\linewidth]{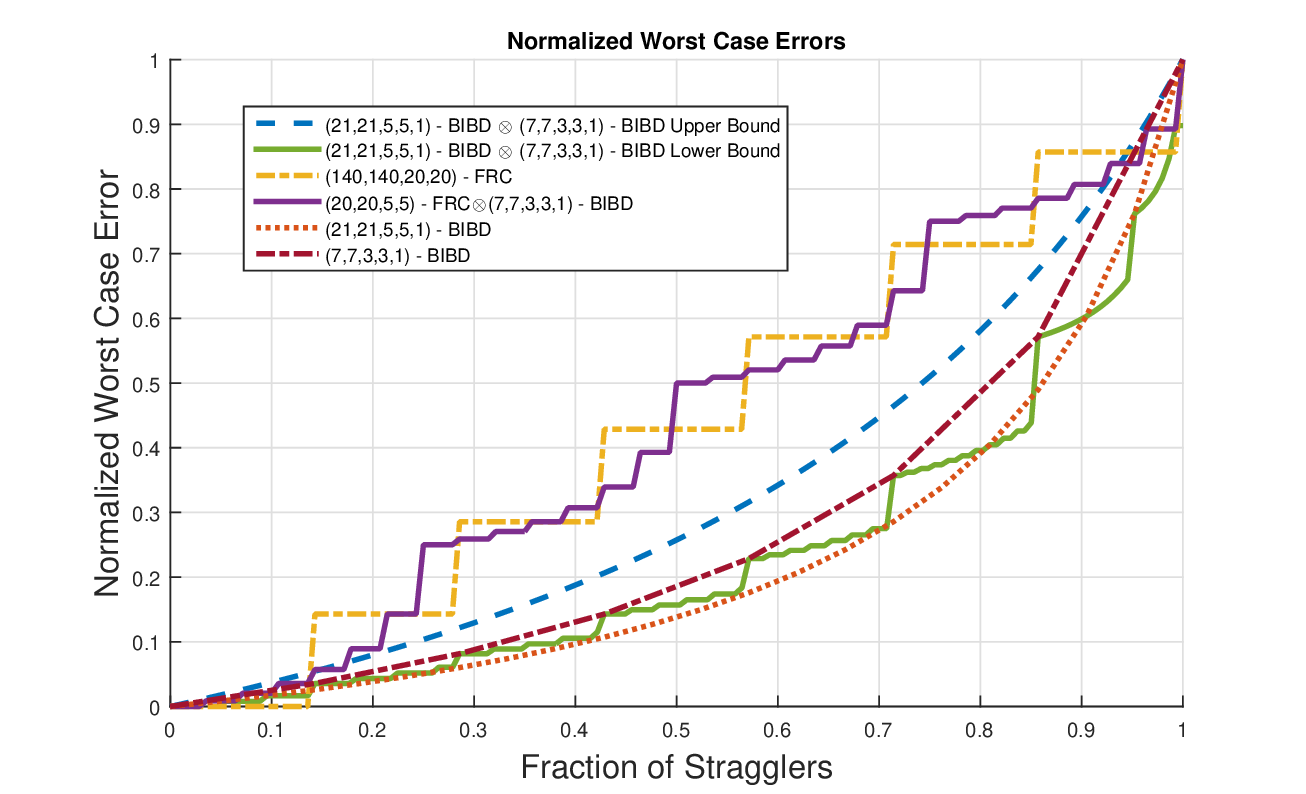}
  \caption{Comparison of Product gradient codes with FRCs and BIBDs.}
  \label{fig:cmp_kron_frc_bibd}
\end{figure}

In Fig~\ref{fig:cmp_kron_frc_bibd}, we plot the errors of two BIBDs and the error bounds on their Kronecker product given by Theorems~\ref{thm:err_kron_bibd} and~\ref{thm:err_kron_bibd_lb}. The figure shows that the kronecker product of BIBDs has comparable performance to the two component BIBDs even though it has a much smaller density.

\subsection{Error of Kronecker Products Involving FRCs}
In this section we prove Theorems~\ref{thm:err_kron_frc} and~\ref{thm:err-frc-bibd}.
\begin{IEEEproof}[\bf Proof of Theorem~\ref{thm:err_kron_frc}]
Notice that $G^\mathrm{F_1} \otimes G^\mathrm{F_2}$ is an $(n_1 n_2, k_1k_2, l_1l_2,r_1r_2)$-FRC. Then the result follows from the error of an FRC.
\end{IEEEproof}

We now establish Theorem~\ref{thm:err-frc-bibd}. In order to prove Theorem~\ref{thm:err-frc-bibd}, we need Lemmas~\ref{lemma:err-general-lambda-gc}--\ref{lemma:convexity_bibd_error}. 
In Lemmas~\ref{lemma:err-general-lambda-gc} and~\ref{lemma:bibd_extension}, we consider an $(n,k,l,r)$ gradient code with exactly $\lambda$ intersections between any pair of workers. We define it as an $(n,k,l,r,\lambda)$-GC.  Notice that an $(n,k,l,r,\lambda)$-GC is not necessarily a BIBD gradient code, since it may not satisfy the BIBD gradient code constraint
\begin{equation*}
    r(l-1) = \lambda (k-1). 
\end{equation*}
However, such a code has the same optimal decoding vector and error expression as a BIBD gradient code with the same parameters $(n,k,l,\lambda)$. 

\begin{lemma}
\label{lemma:err-general-lambda-gc}
Let $G$ be an $(n,k,l,r,\lambda)$-GC, where $l > \lambda$. Then for any set of $s \in [n]$ stragglers $\mathcal{U}$, the optimal decoding vector is given by
\begin{equation*}
    v_\mathrm{opt}(G,\mathcal{U}) = \left( \frac{l}{l + \lambda(n-s-1) }\right) \bm 1_{n-s} ,
\end{equation*}
and the squared error is
\begin{equation*}
    \left\Vert G_\mathcal{U} v_\mathrm{opt}(G,\mathcal{U}) - 1_k \right\Vert_2^2 = k - \frac{l^2(n-s)}{l + \lambda(n-s-1)} .
\end{equation*}
\end{lemma} 

The proof follows essentially from the proof of Theorem 1 in~\cite{Kadhe--Koyluoglu--Ramchandran2019}, and is included in Appendix~\ref{appendix:err-general-lambda-gc} for completeness. 

\begin{lemma}
\label{lemma:bibd_extension}
Let $G^\mathrm{B}$ be an $(n_1,k_1,l_1,\lambda_1)$ gradient code, and $G = G^\mathrm{B}\otimes \bm 1_{k_2}$ for some positive integer $k_2$. Then for any integer $0\le s \le n_1$, and for any $\mathcal{U}$ such that $|\mathcal{U}| = n-s$,
\begin{equation*}
    \left\Vert (G^\mathrm{B}\otimes \bm 1_{k_2})_\mathcal{U} \bm v_\mathrm{opt}(G,\mathcal{U}) - \bm 1_{k_1k_2} \right\Vert_2^2 = k_1k_2 \error(G^\mathrm{B},s). 
\end{equation*}
Equivalently, the normalized worst-case squared errors satisfy
\begin{equation*}
    \error(G^\mathrm{B}\otimes \bm 1_{k_2}, s) = \error(\bm 1_{k_2} \otimes G^\mathrm{B} , s) = \error(G^\mathrm{B},s). 
\end{equation*} 
\end{lemma} 

The proof of Lemma~\ref{lemma:bibd_extension} is deferred to Appendix~\ref{appendix:bibd_extension}.

Another key observation relates the error expressions of any gradient code $G$ and the gradient code formed by taking multiple copies of the workers of $G$, where all copies compute the same set of gradients. The proof is deferred to Appendix~\ref{appendix:gc_replication}.
\begin{lemma}
\label{lemma:gc_replication}
Let $G$ be an $(n,k,l,\lambda)$ gradient code. Then for any $s\in[n]$ and any $r \in \mathbb{N}$,
\begin{equation*}
    \error(\bm 1_{r}^T\otimes G, s) = \error\left( G, \left\lfloor \frac{s}{r} \right\rfloor \right).
\end{equation*}
\end{lemma}

An important observation in establishing  Theorem~\ref{thm:err-frc-bibd} is that the squared error of a block diagonal gradient code is the sum of squared errors from each block.  The following lemma makes this observation precise. The proof is provided in Appendix~\ref{appendix:block_diag_norm}.

\begin{lemma}
\label{lemma:block_diag_norm}
Let $G$ be a block-diagonal gradient code given by 
\begin{equation*}
    G = \begin{bmatrix}
 G^{(1)} &  0_{l\times r} & \dotsc &  0_{l\times r} \\
 0_{l\times r} &  G^{(2)}  & \dotsc &  0_{l\times r} \\
\vdots & \vdots& \ddots & \vdots \\
 0_{l\times r} &  0_{l\times r} & \dotsc & G^{(\tau)}
\end{bmatrix},
\end{equation*}
where $\tau\in\mathbbm{N}$, and each block $G^{(m)}$ is a gradient code with $n$ workers and $k$ gradients to compute for each $m\in[\tau]$. Let $\bm v := (\bm v_1,\bm v_2,\dotsc,\bm v_\tau)$, where $\bm v_m$ is a decoding vector for $G^{(m)}$ for each $m\in[\tau]$. Then for any set of non-stragglers $\mathcal{U}$
\begin{equation*}
    \Vert G_\mathcal{U} \bm v - \bm 1_{k\tau} \Vert_2^2 = \sum_{m=1}^\tau \Vert G^{(m)}_{\mathcal{U}_m} \bm v_m - \bm 1_{k} \Vert_2^2
\end{equation*}
where 
\begin{equation*}
    \mathcal{U}_m = \mathcal{U}\cap\{(m-1)n+1,(m-1)n+2,\dotsc,m n\}
\end{equation*}
is the set of non-stragglers of $G^{(m)}$. 
\end{lemma}

The next key observation in establishing Theorem~\ref{thm:err-frc-bibd} is that the error function of a BIBD is a convex sequence. The proof is deferred to Appendix~\ref{appendix:convexity_bibd_error}. 

\begin{lemma}
\label{lemma:convexity_bibd_error}
Let $G^\mathrm{B}$ be an $(n,k,l,\lambda)$-BIBD. Then $\{\error(G^\mathrm{B},s)\}_{s=0}^n$ is a convex sequence, i.e., 
\[
2\error( G^\mathrm{B},s) \le \error( G^\mathrm{B},s-1)+\error( G^\mathrm{B},s+1)
\]
 for all $s \in[n-1]$.
\end{lemma}

We are ready to  prove Theorem~\ref{thm:err-frc-bibd}. 

\begin{IEEEproof}[\bf Proof of Theorem~\ref{thm:err-frc-bibd}] 
For convenience, let $G := G^\mathrm{F} \otimes G^\mathrm{B}$. Note that the subscript $1$ corresponds to the FRC $G^\mathrm{F}$, and the subscript $2$ corresponds to the BIBD $G^\mathrm{B}$. Recall that for any $s$ stragglers and $\tilde{s} = n_1n_2-s$ non-stragglers, 
\begin{equation*}
    \error(G,s) = \frac{1}{k_1k_2}\max_{\mathcal{U}:|\mathcal{U}| = \tilde{s}} \Vert G_\mathcal{U} \bm v_\mathrm{opt} - \bm 1_{k_1k_2} \Vert_2^2. 
\end{equation*}
We simplify the squared error $\Vert G_\mathcal{U} \bm v - \bm 1_{k_1k_2} \Vert_2^2$ for any set $\tilde{s}$ non-stragglers $\mathcal{U}$ by decomposing $G$ into blocks. Then, we simplify the resulting expression to determine $\error(G,s)$.

By construction, $G$ is a block-diagonal matrix given by 
\begin{equation*}
    G = 
    \begin{bmatrix}
     J_{l_1\times r_1} \otimes G^\mathrm{B} &  {\bm 0} & \dotsc &  {\bm 0} \\
     \bm 0 &  J_{l_1\times r_1} \otimes G^\mathrm{B}  & \dotsc &  \bm 0 \\
    \vdots & \vdots& \ddots & \vdots \\
     \bm 0 &  \bm 0 & \dotsc & J_{l_1\times r_1} \otimes G^\mathrm{B}
\end{bmatrix},
\end{equation*}
where each zero block $\bm 0$ is the $l_1 k_2 \times r_1 n_2$ all-zero matrix. For each block $m\in [k_1/l_1]$, let 
$
    \mathcal{U}_m = \mathcal{U} \cap  \{ (m-1)n_2r_1 + 1, (m-1)n_2r_1 + 1,\dotsc, m n_2r_1 \}
$
be the set of non-stragglers in block $m$. Denote $\tilde{s}_m = \left | \mathcal{U}_m \right| $ as the number of non-stragglers and $s_m = n_1r_2 - \tilde{s}_m$ as the number of straggling workers in block $m$. 

Now, represent the optimal decoding vector in the following form 
\begin{equation*}
    \bm v_\mathrm{opt} = (\bm v_1^T, \bm v_2^T,\dotsc, \bm v_{k_1/l_1}^T)^T, 
\end{equation*}
where $\bm v_m$ is a length $\tilde{s}_m$ vector for each block $m$. Then 
\begin{align*}
    &\hspace{1.25em} \Vert G_\mathcal{U} \bm v_\mathrm{opt} - \bm 1_{k_1k_2} \Vert_2^2 \\
    &\stackrel{(a)}{=} \sum_{m=1}^{k_1/l_1} \big\Vert \left(J_{l_1\times r_1} \otimes G^\mathrm{B} \right)_{\mathcal{U}_m} \bm v_m - \bm 1_{l_1k_2} \big\Vert_2^2  \\
    &= \sum_{m=1}^{k_1/l_1} \big\Vert \left(\left(\bm 1_{l_1} \otimes \bm 1_{r_1}^T \right) \otimes G^\mathrm{B} \right)_{\mathcal{U}_m} \bm v_m - \bm 1_{l_1k_2} \big\Vert_2^2  \\
    &= \sum_{m=1}^{k_1/l_1} \big\Vert \left(\bm 1_{l_1} \otimes \left(\bm 1_{r_1}^T  \otimes G^\mathrm{B} \right)\right)_{\mathcal{U}_m} \bm v_m - \bm 1_{l_1k_2} \big\Vert_2^2  \\
    &\stackrel{(b)}{=} \sum_{m=1}^{k_1/l_1}  l_1 k_2 \error \left(\bm 1_{r_1}^T  \otimes G^\mathrm{B} , s_m \right) \\
    &\stackrel{(c)}{=} \sum_{m=1}^{k_1/l_1}  l_1 k_2 \error\left(G^\mathrm{B}, \left\lfloor \tfrac{s_m}{r_1} \right\rfloor\right),
\end{align*}
where $(a)$ follows from Lemma~\ref{lemma:block_diag_norm}, $(b)$ follows from Lemma~\ref{lemma:bibd_extension}, and $(c)$ follows from Lemma~\ref{lemma:gc_replication}.

We now have
\begin{align*}
    \error(G,s)
    &= \frac{1}{k_1k_2}\max_{\mathcal{U}:|\mathcal{U}| = \tilde{s}} \sum_{m=1}^{k_1/l_1}  l_1 k_2 \error\left(G^\mathrm{B}, \left\lfloor \frac{s_m}{r_1} \right\rfloor\right) \\
    &\stackrel{(a)}{=} \frac{1}{k_1k_2} \Bigg( \sum_{m=1}^c l_1k_2 \error \left(G^\mathrm{B}, \left\lfloor \frac{s_m}{r_1} \right\rfloor \right) + l_1k_2 \error \left(G^\mathrm{B}, \left\lfloor \frac{s_{c+1}}{r_1} \right\rfloor \right) \Bigg)\\
    &\stackrel{(b)}{=} \frac{l_1}{k_1} \left( \left\lfloor \frac{s}{r_1n_2} \right\rfloor + \error\left(G^\mathrm{B},b \right) \right) \\
    &= \error\left( G^\mathrm{F}, \frac{s}{n_2} \right) + \frac{l_1}{k_1} \error\left(G^\mathrm{B},b \right).
\end{align*}
Above, $(a)$ follows since the error expression of $G^\mathrm{B}$ is convex as shown in Lemma~\ref{lemma:convexity_bibd_error}, therefore
\begin{equation*}
    \sum_{m=1}^{k_1/l_1} l_1 k_2 \error \left(G^\mathrm{B}, \left\lfloor \frac{s_m}{r_1} \right\rfloor \right)
\end{equation*} 
is maximized when all workers in $c := \left\lfloor \tfrac{s}{n_2 r_1} \right\rfloor$
blocks are straggled, i.e., $s_1 = s_2 = \dotsc = s_c = r_1n_2$, and the remaining stragglers are placed in block $c+1$. Moreover, $(b)$ follows by observing that the first $c$ blocks each have $r_1 n_2$ stragglers and $\error(G^\mathrm{B}, \lfloor (r_1 n_2)/r_1 \rfloor ) = 1$.
\end{IEEEproof}

\begin{remark}
Observe that in the proof of Theorem~\ref{thm:err-frc-bibd}, the only property of the BIBD gradient code we use is the convexity of the BIBD gradient code error expression. Thus, the result in Theorem~\ref{thm:err-frc-bibd} generalizes to any gradient code with an error expression that is convex in the number of stragglers. 
\end{remark}

\subsection{Error of Kronecker Products of BIBDs}
In this section, we prove Theorems~\ref{thm:err_kron_bibd} and~\ref{thm:err_kron_bibd_lb}, which give upper and lower bounds respectively on the error of Kronecker products of BIBDs. 

To establish Theorem~\ref{thm:err_kron_bibd}, which upper bounds the normalized worst-case squared error of the Kronecker product of two BIBDs, we need the following technical lemma (see  Appendix~\ref{appendix:err_kron_bibd} for the proof).

\begin{lemma}
\label{lemma:err_kron_bibd}
Let $G^\mathrm{B_i}$ be an $(n_i,k_i,l_i,r_i,\lambda_i)$-BIBD for $i=1,2$, and $G=G^\mathrm{B_1}\otimes G^\mathrm{B_2}$. Then for any set of $\tilde{s}$ non-stragglers $\mathcal{U}\subset [n_1n_2]$,
\begin{equation*}
    \bm 1_{\tilde{s}}^T G_\mathcal{U}^T  G_\mathcal{U} \bm 1_{\tilde{s}} \le \tilde{s}d + \lambda_1\lambda_2 \tilde{s}^2,
\end{equation*}
where 
\begin{equation*}
    d = (l_1-\lambda_1)(l_2-\lambda_2)+n_2(l_1\lambda_2-\lambda_1\lambda_2) + n_1(\lambda_1l_2-\lambda_1\lambda_2).
\end{equation*}
\end{lemma} 

Notice that the Kronecker product of two BIBD gradient codes is not necessarily another BIBD gradient code. As a result, a constant decoding vector may not be optimal. However, sub-optimal constant decoding vectors provide reasonable error performance in simulations. In the following, we establish Theorem~\ref{thm:err_kron_bibd} by upper bounding the error using the error corresponding to a constant decoding vector. 

\begin{IEEEproof}[\bf Proof of Theorem~\ref{thm:err_kron_bibd}]
For convenience we write $G:=G^\mathrm{B_1} \otimes G^\mathrm{B_2}$. Consider a constant decoding vector $\bm v_a$ given by
\begin{equation*}
    \bm v_a = a \bm 1_{\tilde{s}},
\end{equation*}
where $a\in\mathbbm{R}$ will be specified later, and $\tilde{s}:= n_1n_2-s$ is the number of non-straggling workers. Then
\begin{align*}
    \error(G,s)
    &= \frac{1}{k_1k_2} \max_{\mathcal{U}: |\mathcal{U}|=\tilde{s}} \min_{\bm v} \Vert G_\mathcal{U} \bm v - \bm 1_{k_1k_2} \Vert_2^2 \\
    &\le \frac{1}{k_1k_2} \max_{\mathcal{U}: |\mathcal{U}|=\tilde{s}} \min_{a} \Vert G_\mathcal{U} \bm v_a - \bm 1_{k_1k_2} \Vert_2^2 \\
    &\stackrel{(a)}{=} \frac{1}{k_1k_2} \max_{\mathcal{U}: |\mathcal{U}|=\tilde{s}} \min_{a} \left( k_1k_2 - 2 \bm v_a^T G_\mathcal{U}^T \bm 1_{k_1k_2} + \bm v_a^T G_\mathcal{U}^T  G_\mathcal{U} \bm v_a \right) \\
    &\stackrel{(b)}{\le} \frac{1}{k_1k_2} \max_{\mathcal{U}: |\mathcal{U}|=\tilde{s}} \min_{a} \Big( k_1k_2 -2a l_1l_2\tilde{s} + a^2 \left(\tilde{s}d + \lambda_1\lambda_2 \tilde{s}^2 \right) \Big) \\
    &= \frac{1}{k_1k_2} \min_{a} \Big( k_1k_2 -2a l_1l_2\tilde{s} + a^2 \left (\tilde{s}d + \lambda_1\lambda_2 \tilde{s}^2 \right) \Big) \\
    &\stackrel{(c)}{=} 1 - \frac{(l_1l_2)^2 (\tilde{s})}{k_1k_2(d + \lambda_1\lambda_2\tilde{s})}.
\end{align*}
Firstly, $(a)$ follows since for any set of $\tilde{s}$ non-stragglers $\mathcal{U}$, we have
\begin{align*}
    &\left\Vert G_\mathcal{U}\bm v_a - \bm 1_{k_1k_2} \right\Vert_2^2 \\
    =& \left( G_\mathcal{U}\bm v_a - \bm 1_{k_1k_2} \right)^T \left( G_\mathcal{U}\bm v_a - \bm 1_{k_1k_2} \right) \\
    =& k_1k_2 - 2 \bm v_a^T G_\mathcal{U}^T \bm 1_{k_1k_2} + \bm v_a^T G_\mathcal{U}^T  G_\mathcal{U} \bm v_a. 
\end{align*}
Secondly, $(b)$ follows since by construction, each column of $G_\mathcal{U}$ has exactly $l_1l_2$ ones, thus
\begin{equation*}
    2 \bm v_a^T G_\mathcal{U}^T \bm 1_{k_1k_2} = 2 a \bm 1_{\tilde{s}}^T  G_\mathcal{U}^T \bm 1_{k_1k_2} 
    = 2 a l_1l_2 \bm 1_{\tilde{s}}^T \bm 1_{\tilde{s}} 
    = 2 a l_1l_2 \tilde{s}. 
\end{equation*}
Moreover, from Lemma~\ref{lemma:err_kron_bibd}, we have
\begin{equation*}
    \bm v_a^T G_\mathcal{U}^T  G_\mathcal{U} \bm v_a 
    = a^2 \bm 1_{\tilde{s}}^T G_\mathcal{U}^T  G_\mathcal{U} \bm 1_{\tilde{s}} 
    \le a^2 \left( \tilde{s}d + \lambda_1\lambda_2 \tilde{s}^2 \right). 
\end{equation*}
Lastly, $(c)$ follows by observing that $l_1\ge\lambda_1$, $l_2\ge\lambda_2$, and $\tilde{s}\ge 0$,  thus we are minimizing over a convex quadratic function in $a$, where the minimizing constant $a^*$ is given by
\begin{equation*}
    a^* = \frac{l_1l_2}{d+\tilde{s}\lambda_1\lambda_2} = \frac{l_1l_2}{d+(n_1n_2-s)\lambda_1\lambda_2}. 
\end{equation*} 
\end{IEEEproof} 

\begin{IEEEproof}[\bf Proof of Theorem~\ref{thm:err_kron_bibd_lb}]
For convenience we write $G:=G^\mathrm{B_1} \otimes G^\mathrm{B_2}$. To establish the result, we consider a set of $n_1n_2-s$ non-stragglers $\mathcal{U}$ satisfying 
\begin{equation*}
    G_\mathcal{U} = G^\mathrm{B_1}_{\mathcal{U}_1} \otimes G^\mathrm{B_2}_{\mathcal{U}_2},
\end{equation*}
where $\mathcal{U}_1$ and $\mathcal{U}_2$ are sets of $n_1 -s_1$ and $n_2 - s_2$ non stragglers respectively, satisfying $(n_1n_2 - s) = (n_1 -s_1)(n_2 - s_2)$ and $\mathcal{U} = \{n_2(u_1 - 1) + u_2 | u_1\in\mathcal{U}_1, u_2\in\mathcal{U}_2\}$.
Since $\mathcal{U}$ is not necessarily the worst-case set of non-stragglers, we have 
\begin{equation*}
    \error(G^\mathrm{B_1} \otimes G^\mathrm{B_2}, s) \ge \frac{1}{k_1k_2}\Vert G_\mathcal{U} \bm v_\mathrm{opt}(G,\mathcal{U}) - \bm 1_{k_1k_2} \Vert_2^2.
\end{equation*}
Notice that 
\begin{align*}
    v_\mathrm{opt}(G,\mathcal{U})
    &= G_\mathcal{U}^\dagger \bm 1_{k_1k_2} \\
    &= (G^\mathrm{B_1}_{\mathcal{U}_1} \otimes G^\mathrm{B_2}_{\mathcal{U}_2})^\dagger \bm 1_{k_1k_2} \\
    &\stackrel{(a)}{=} \left(\left( G^\mathrm{B_1}_{\mathcal{U}_1} \right)^\dagger \otimes \left( G^\mathrm{B_2}_{\mathcal{U}_2} \right)^\dagger \right) \bm 1_{k_1k_2} \\
    &= \left(\left( G^\mathrm{B_1}_{\mathcal{U}_1} \right)^\dagger \otimes \left( G^\mathrm{B_2}_{\mathcal{U}_2} \right)^\dagger \right) \left(\bm 1_{k_1} \otimes \bm 1_{k_2} \right) \\
    &\stackrel{(b)}{=} \left( \left( G^\mathrm{B_1}_{\mathcal{U}_1} \right)^\dagger  \bm 1_{k_1} \right) \otimes \left( \left( G^\mathrm{B_2}_{\mathcal{U}_2} \right)^\dagger  \bm 1_{k_2} \right) \\
    &= \bm v_\mathrm{opt}(G^\mathrm{B_1}, \mathcal{U}_1) \otimes \bm v_\mathrm{opt}(G^\mathrm{B_2}, \mathcal{U}_2),
\end{align*}
where $(a)$ follows from~\cite[eqn. (222)]{Petersen-Pedersen2012}, and $(b)$ follows from the mixed product property~\cite[eqn. (511)]{Petersen-Pedersen2012} of Kronecker products. 
Then observe that 
\begin{equation*}
    \Vert G_\mathcal{U} \bm v_\mathrm{opt}(G,\mathcal{U}) - \bm 1_{k_1k_2} \Vert_2^2 
    = k_1k_2 - 2 \bm v_\mathrm{opt}(G,\mathcal{U})^T G_\mathcal{U}^T \bm 1_{k_1k_2} + \bm v_\mathrm{opt}(G,\mathcal{U})^T G_\mathcal{U}^T G_\mathcal{U} \bm v_\mathrm{opt}(G,\mathcal{U}).
\end{equation*}
We have
\begin{align*}
    &\phantom{ = }2 \bm v_\mathrm{opt}(G,\mathcal{U})^T G_\mathcal{U}^T \bm 1_{k_1k_2} \\
    &= 2 \left( \bm v_\mathrm{opt}(G^\mathrm{B_1}, \mathcal{U}_1) \otimes \bm v_\mathrm{opt}(G^\mathrm{B_2}, \mathcal{U}_2) \right)^T (G^\mathrm{B_1}_{\mathcal{U}_1} \otimes G^\mathrm{B_2}_{\mathcal{U}_2})^T \left(\bm 1_{k_1} \otimes 1_{k_2} \right) \\
    &\stackrel{(a)}{=} 2 \left( \bm v_\mathrm{opt}(G^\mathrm{B_1}, \mathcal{U}_1)^T \left(G^\mathrm{B_1}_{\mathcal{U}_1} \right)^T \bm 1_{k_1} \right) \left( \bm v_\mathrm{opt}(G^\mathrm{B_2}, \mathcal{U}_2)^T \left(G^\mathrm{B_2}_{\mathcal{U}_2} \right)^T \bm 1_{k_2} \right) \\ 
    &\stackrel{(b)}{=} 2c_1 c_2.
\end{align*}
Above, $(a)$ follows from the mixed product property~\cite[eqn. (511)]{Petersen-Pedersen2012} of Kronecker products, and $(b)$ follows from~\cite[eqn. (17)]{Kadhe--Koyluoglu--Ramchandran2019}.
Similarly, 
\begin{align*}
    &\phantom{ = }\bm v_\mathrm{opt}(G,\mathcal{U})^T G_\mathcal{U}^T G_\mathcal{U} \bm v_\mathrm{opt}(G,\mathcal{U}) \\
    &= \left( \bm v_\mathrm{opt}(G^\mathrm{B_1}, \mathcal{U}_1) \otimes \bm v_\mathrm{opt}(G^\mathrm{B_2}, \mathcal{U}_2) \right)^T (G^\mathrm{B_1}_{\mathcal{U}_1} \otimes G^\mathrm{B_2}_{\mathcal{U}_2})^T (G^\mathrm{B_1}_{\mathcal{U}_1} \otimes G^\mathrm{B_2}_{\mathcal{U}_2}) \left( \bm v_\mathrm{opt}(G^\mathrm{B_1}, \mathcal{U}_1) \otimes \bm v_\mathrm{opt}(G^\mathrm{B_2}, \mathcal{U}_2) \right) \\
    &= \left( \bm v_\mathrm{opt}(G^\mathrm{B_1}, \mathcal{U}_1)^T \left(G^\mathrm{B_1}_{\mathcal{U}_1} \right)^T \left(G^\mathrm{B_1}_{\mathcal{U}_1} \right) \bm v_\mathrm{opt}(G^\mathrm{B_1}, \mathcal{U}_1)  \right) \otimes \left( \bm v_\mathrm{opt}(G^\mathrm{B_2}, \mathcal{U}_2)^T \left(G^\mathrm{B_2}_{\mathcal{U}_2} \right)^T \left(G^\mathrm{B_2}_{\mathcal{U}_2} \right) \bm v_\mathrm{opt}(G^\mathrm{B_2}, \mathcal{U}_2)  \right) \\
    &\stackrel{(a)}{=} d_1 d_2,
\end{align*}
where $(a)$ follows from~\cite[eqn. (17)]{Kadhe--Koyluoglu--Ramchandran2019}.
The result follows. 
\end{IEEEproof}

\begin{remark}
Let $G:=G^\mathrm{(1)} \otimes G^\mathrm{(2)}$ be any gradient code where $G^\mathrm{(1)}$ and $G^\mathrm{(2)}$ can be realizations of probabilistic constructions, and let $\mathcal{U}$ be any set of $n_1n_2-s$ non-stragglers  satisfying 
\begin{equation*}
    G_\mathcal{U} = G^\mathrm{(1)}_{\mathcal{U}_1} \otimes G^\mathrm{(2)}_{\mathcal{U}_2},
\end{equation*}
where $\mathcal{U}_1$ and $\mathcal{U}_2$ are sets of $n_1 -s_1$ and $n_2 - s_2$ non stragglers respectively, satisfying $(n_1n_2 - s) = (n_1 -s_1)(n_2 - s_2)$ and $\mathcal{U} = \{n_2(u_1 - 1) + u_2 | u_1\in\mathcal{U}_1, u_2\in\mathcal{U}_2\}$.
Then we have
\begin{equation*}
     v_\mathrm{opt}(G,\mathcal{U})
    = \bm v_\mathrm{opt}(G^\mathrm{(1)}, \mathcal{U}_1) \otimes \bm v_\mathrm{opt}(G^\mathrm{(2)}, \mathcal{U}_2)
\end{equation*}
from the properties of Kronecker products as shown
in the proof of Theorem~\ref{thm:err_kron_bibd_lb}.
\end{remark}

\subsection{Error of Kronecker Products of Soft BIBDs}

To establish Theorem~\ref{thm:err_kron_prob_bibd}, we make use of a sub-optimal constant decoding vector as done in the proof of Theorem~\ref{thm:err_kron_bibd}. The proof of Theorem~\ref{thm:err_kron_prob_bibd} is similar to the proof of Theorem~\ref{thm:err_kron_bibd}.

\begin{IEEEproof}[\bf Proof of Theorem~\ref{thm:err_kron_prob_bibd}]
For convenience, we write $\bm G := \bm G^\mathrm{PB1} \otimes \bm G^\mathrm{PB2}$. We consider the constant decoding vector $\bm v_a = a \bm 1_{\tilde{s}}$, where $a \in \mathbb{R}$ will be specified later and $\tilde{s} := n_1n_2 - s$ is the number of non-straggling workers. Then
\begin{align*}
    &\hspace{1.5em} \error(\bm G,s) \\
    &= \frac{1}{k_1k_2} \max_{\mathcal{U}: |\mathcal{U}|=\tilde{s}} \mathbb{E} \left[ \min_{\bm v} \Vert \bm G_\mathcal{U} \bm v - \bm 1_{k_1k_2} \Vert_2^2 \right] \\
    &\stackrel{}{\le} \frac{1}{k_1k_2} \max_{\mathcal{U}: |\mathcal{U}|=\tilde{s}} \mathbb{E} \left[ \min_{a} \Vert \bm G_\mathcal{U} \bm v_a - \bm 1_{k_1k_2} \Vert_2^2 \right] \\
    &\stackrel{}{\le} \frac{1}{k_1k_2} \max_{\mathcal{U}: |\mathcal{U}|=\tilde{s}} \min_{a} \mathbb{E} \left[\Vert \bm G_\mathcal{U} \bm v_a - \bm 1_{k_1k_2} \Vert_2^2 \right] \\
    &\stackrel{}{=} \frac{1}{k_1k_2} \max_{\mathcal{U}: |\mathcal{U}|=\tilde{s}} \min_{a} \mathbb{E}\left[ k_1k_2 - 2 \bm v_a^T \bm G_\mathcal{U}^T \bm 1_{k_1k_2} + \bm v_a^T \bm G_\mathcal{U}^T  \bm G_\mathcal{U} \bm v_a \right] \\
    &\stackrel{(a)}{\le} \frac{1}{k_1k_2} \max_{\mathcal{U}: |\mathcal{U}|=\tilde{s}} \min_{a} \Big( k_1k_2 -2a l_1l_2\tilde{s} + \mathbb{E} \left[ \bm v_a^T \bm G_\mathcal{U}^T  \bm G_\mathcal{U} \bm v_a \right] \Big) \\
    &\stackrel{(b)}{\le} \frac{1}{k_1k_2} \max_{\mathcal{U}: |\mathcal{U}|=\tilde{s}} \min_{a} \Big( k_1k_2 -2a l_1l_2\tilde{s} + a^2 \left(\tilde{s}d + \lambda_1\lambda_2 \tilde{s}^2 \right) \Big) \\
    &\stackrel{}{=} 1 - \frac{(l_1l_2)^2 (\tilde{s})}{k_1k_2(d + \lambda_1\lambda_2\tilde{s})}.
\end{align*}
Above, $(a)$ follows by the linearity of expectation, and from the fact that on average each column of $\bm G$ has $l_1l_2$ ones entries and $k_1k_2 - l_1l_2$ zero entries i.e. $\mathbb{E}\left[ \bm G_\mathcal{U}^T \bm 1_{k_1k_2} \right] = l_1l_2 \bm 1_{\tilde{s}}$. Therefore
\begin{equation*}
    \mathbb{E}\left[ 2 \bm v_a^T \bm G_\mathcal{U}^T \bm 1_{k_1k_2}  \right] 
    = 2 \bm v_a^T \mathbb{E}\left[ \bm G_\mathcal{U}^T \bm 1_{k_1k_2} \right] 
    = 2 \bm a \bm 1_{\tilde{s}}^T \left( l_1l_2 \bm 1_{\tilde{s}} \right) 
    = 2 a l_1 l_2 \tilde{s}.
\end{equation*}
Moreover, $(b)$ follows since we have
\begin{equation}
    \label{eqn:pbibd_worker_dot_prod}
    \mathbb{E} \left[ \bm v_a^T \bm G_\mathcal{U}^T  \bm G_\mathcal{U} \bm v_a \right] 
    \leq a^2 \left( \tilde{s}d + \lambda_1\lambda_2 \tilde{s}^2 \right).
\end{equation}
To show that~\eqref{eqn:pbibd_worker_dot_prod} holds, we first write 
\begin{equation*}
    \bm G_\mathcal{U} = 
    \begin{bmatrix}
    \vert & \vert & \dotsc & \vert \\
    \bm w_1 & \bm w_2 & \dotsc &  \bm w_{\tilde{s}} \\
    \vert & \vert & \dotsc & \vert 
\end{bmatrix},
\end{equation*}
where $\bm w_1, \bm w_2,\dotsc, \bm w_{\tilde{s}}$ are random column vectors of size $k_1k_2$. Then 
\begin{equation*}
    \mathbb{E} \left[ \bm v_a^T \bm G_\mathcal{U}^T  \bm G_\mathcal{U} \bm v_a \right]
    = a^2 \mathbb{E} \left[ \bm 1_{\tilde{s}}^T \bm G_\mathcal{U}^T  \bm G_\mathcal{U} \bm 1_{\tilde{s}} \right] 
    = a^2 \mathbb{E} \left[ \sum_{i\in\mathcal{U}} \sum_{i\in\mathcal{U}} \bm w_i^T \bm w_j \right] 
    = a^2  \sum_{i\in\mathcal{U}} \sum_{i\in\mathcal{U}} \mathbb{E} \left[ \bm w_i^T \bm w_j \right ].
\end{equation*}
We now introduce some notation to help simplify the $\mathbb{E} \left[ \bm w_i^T \bm w_j \right ]$ term. Let block $m\in[n_1]$ refer to the set of $n_2$ workers 
\begin{equation*}
    \mathcal{B}_m := \{(m-1)n_2+1,(m-1)n_2+2,\dotsc,m n_2\}. 
\end{equation*}
Similarly, let class $m'\in[n_1]$ refer to the set of $n_1$ workers
\begin{equation*}
    \mathcal{C}_{m'} := \{m', m'+n_2, m'+2n_2,\dotsc, m'+{n_1-1}n_2\}.
\end{equation*}
Let $i_\mathrm{b}$ and $j_\mathrm{b}$ be the block  that $i$ and $j$ are from respectively and let $i_\mathrm{c}$ and $j_\mathrm{c}$ be the classes that workers $i$ and $j$ are in. Let $\bm w_{i,z}$ denote entry $z$ of the random vector $\bm w_i$. Lastly let $p_1(x^{n_1})$ and $p_2(x^{n_2})$ be the distributions according to which the rows of $\bm G^\mathrm{PB1}$ and $\bm G^\mathrm{PB1}$ are generated from respectively. 
Then
\begin{align}
    \nonumber
    \mathbb{E} \left[ \bm w_i^T \bm w_j \right ] %
    &= \mathbb{E} \left[ \sum_{z=1}^{k_1k_2}  \bm w_{i,z} \bm w_{j,z} \right] \\ \nonumber
    &= \mathbb{E} \Bigg[ \sum_{t=1}^{k_1} \bm G_{t, i_b}^\mathrm{PB1} \bm G_{t, j_b}^\mathrm{PB1} ( \bm w_{i,1}\bm w_{j,1} + \bm w_{i,2}\bm w_{j,2}  + \dotsc + \bm w_{i,k_2}\bm w_{j,k_2} ) \Bigg] \\ \nonumber
    &= \mathbb{E} \left[ \left( \sum_{t=1}^{k_1} \bm G_{t, i_b}^\mathrm{PB1} \bm G_{t, j_b}^\mathrm{PB1} \right) \left( \sum_{t=1}^{k_2} \bm G_{t, i_c}^\mathrm{PB2} \bm G_{t, j_c}^\mathrm{PB2} \right) \right] \\ \nonumber
    &\stackrel{(b)}{=}  \mathbb{E}_{p_1} \left[  \sum_{t=1}^{k_1} \bm G_{t, i_b}^\mathrm{PB1} \bm G_{t, j_b}^\mathrm{PB1} \right] \mathbb{E}_{p_2} \left[ \sum_{t=1}^{k_2} \bm G_{t, i_c}^\mathrm{PB2} \bm G_{t, j_c}^\mathrm{PB2} \right] \\ 
    \label{eqn:pbibd_cases}
    &\stackrel{(c)}{=} \begin{cases} 
      l_1l_2 & i=j, \\ 
      l_1\lambda_2 & i\neq j, i_\mathrm{b} = j_\mathrm{b}, \\ 
      \lambda_1l_2 & i\neq j, i_\mathrm{c} = j_\mathrm{c}, \\ 
      \lambda_1\lambda_2 & \text{otherwise.}
   \end{cases}
\end{align}
Above, $(b)$ follows since $\bm G^\mathrm{PB1}$ and $\bm G^\mathrm{PB2}$ are generated independently from each other according to distributions $p_1(x^{n_1})$ and $p_2(x^{n_2})$ respectively. Furthermore, $(c)$ follows from Lemma~\ref{lemma:err_kron_bibd} and by observing that on average $\bm G^\mathrm{PB1}$ and $\bm G^\mathrm{PB2}$ have the same structures as combinatorial $(n_1,k_1,l_1,\lambda_1)$ and $(n_1,k_1,l_1,\lambda_1)$ gradient codes respectively. 
Then 
\begin{equation*}
    \mathbb{E} \left[ \bm v_a^T \bm G_\mathcal{U}^T  \bm G_\mathcal{U} \bm v_a \right] = a^2  \sum_{i\in\mathcal{U}} \sum_{i\in\mathcal{U}} \mathbb{E} \left[ \bm w_i^T \bm w_j \right ] 
    \stackrel{(d)}{\leq} a^2 \left( \tilde{s}d + \lambda_1\lambda_2 \tilde{s}^2 \right),
\end{equation*}
where $(d)$ follows from equation~\eqref{eqn:pbibd_cases} and Lemma~\ref{lemma:err_kron_bibd}. 
\end{IEEEproof}

\section{Conclusion} 
In this work, we provide two approximate gradient code constructions. We propose Soft BIBD gradient codes which allows us to construct gradient codes for a wider range of system parameters than BIBD gradient codes, and has smaller average squared error than BIBD gradient codes. Our second construction, called product gradient codes, allows us to construct new gradient codes from existing gradient codes in a scalable manner. We derive upper bounds and lower bounds on the normalized worst-case squared errors of various product gradient codes. Fig.~\ref{fig:cmp_kron_frc_bibd} shows that the Kronecker product of BIBDs has comparable error performance to the component BIBD gradient codes even though it has much smaller density.

\bibliographystyle{IEEEtran}
\bibliography{references_arxiv}

\newpage
\appendices

\section{FRC Error}
\label{appendix:error_frc}
\begin{lemma}
\label{lemma:error_frc}
If $G^{\mathrm{F}}$ is an $(n,k,l,r)$-FRC, then the error under $s$ stragglers is 
\begin{equation*}
\error(G^{\mathrm{F}},s) = \frac{l}{k} \left \lfloor  \frac{s}{r} \right \rfloor
\end{equation*}
for any $s\in[n_1]$.
\end{lemma}

\begin{IEEEproof}[\bf Proof]
Recall that
\begin{equation*}
G^\mathrm{F} = 
\begin{bmatrix}
 J_{l\times r} &  0_{l\times r} & \dotsc &  0_{l\times r} \\
 0_{l\times r} &  J_{l\times r}  & \dotsc &  0_{l\times r} \\
\vdots & \vdots& \ddots & \vdots \\
 0_{l\times r} &  0_{l\times r} & \dotsc & J_{l\times r}
\end{bmatrix}.
\end{equation*}
Therefore
\begin{align*}
    \error(G^{\mathrm{F}},s) &= \max_{\mathcal{U}: |\mathcal{U}|=n-s} \frac{1}{k}\Vert  G^\mathrm{F}_\mathcal{U} \bm v_\mathrm{opt}(G^\mathrm{F}, \mathcal{U}) - \bm 1_k \Vert_2^2 \\
    &\stackrel{(a)}{=} \max_{\mathcal{U}: |\mathcal{U}|=n-s} \frac{1}{k} \sum_{m=1}^{k/l} \Vert  (J_{l\times r})_{\mathcal{U}_m} \bm v_\mathrm{opt}(J_{l\times r}, \mathcal{U}_m) - \bm 1_l \Vert_2^2 \\
    &\stackrel{(b)}{=} \frac{1}{k} \sum_{m=1}^{\lfloor s/r \rfloor} \Vert  \bm 1_l \Vert_2^2 \\
    &= \frac{l}{k} \left \lfloor  \frac{s}{r} \right \rfloor,
\end{align*}
where $(a)$ follows from Lemma~\ref{lemma:block_diag_norm}, and $(b)$ follows since the all-ones gradient code with encoding matrix $J_{l\times r}$ has zero error when there are less than $r$ stragglers, and cannot recover any of the $l$ gradients when there are $r$ stragglers. Thus, in the worst case, $\lfloor  \tfrac{s}{r}  \rfloor$ copies of $J_{l\times r}$ are fully straggled, while the remaining copies have no stragglers. 
\end{IEEEproof}

\section{Proof of Lemma~\ref{lemma:permutation_equivalence}}
\label{appendix:permutation_equivalence}
Observe that there exist permutation matrices $P$ and $Q$ such that 
\begin{equation*}
    G^{(2)} \otimes G^{(1)} = P \left( G^{(1)} \otimes G^{(2)} \right) Q
\end{equation*}
 
We now define \emph{zeroing} matrices, which will help simplify the error expression. Firstly, we denote by $\bm e_j \in \mathbb{R}^n$ the standard basis vector which has a one in the j-th entry, and zeroes in all other entries. We call the $n \times n$ matrix $T$ a \emph{zeroing} matrix if it has $n-s$ columns which are all distinct standard basis vectors, and $s$ all zeroes columns. 

Let $\mathcal{U} \subset [n_1 n_2]$ be any set of $n_1n_2-s$ non-stragglers. Write 
\begin{equation*}
    Q = [\bm e_{\pi(1)}, \bm e_{\pi(2)}, \dotsc, \bm e_{\pi(n)}]
\end{equation*}
where $\pi$ is some permutation on $\{1,\dotsc,n\}$, and $e_i$ is the $i$-th standard basis vector of length $n$, for each $i\in[n]$. Now define 
\begin{equation*}
    X = [\bm e_{\pi(1)} \mathbbm{1}(\pi(1) \in \mathcal{U}), \bm e_{\pi(2)}\mathbbm{1}(\pi(2) \in \mathcal{U}), \dotsc , \bm e_{\pi(n)}\mathbbm{1}(\pi(n) \in \mathcal{U})].
\end{equation*}
Observe that $X$ is a zeroing matrix. Writing 
${\bm v_Q := \bm v_{opt}(GQ, \mathcal{U})}$, 
we claim that for any gradient code $G$,
\begin{align}
    \nonumber
    &\phantom{=}\Vert (GX (X^T \bm v_{opt}(GQ, \mathcal{U})) - \bm 1_{k_1k_2} \Vert_2^2 \\
    \nonumber
    &=\Vert (G X) (X^T \bm v_Q) - \bm 1_{k_1k_2} \Vert_2^2 \\
    \nonumber
    &= \sum_{i=1}^k \sum_{j=1}^n \left[ (GX)_{ij}(X^T \bm v_Q)_j - \bm 1_{k_1k_2}  \right]^2 \\
    \nonumber
    &= \sum_{i=1}^k \sum_{j=1}^n \left[ (GX)_{ij}(X^T \bm v_Q)_j \mathbbm{1}(j \in \mathcal{U}) - \bm 1_{k_1k_2}  \right]^2 \\
    \nonumber
    &= \sum_{i=1}^k \sum_{j\in\mathcal{U}} \left[ (GX)_{ij}(X^T \bm v_Q)_j - \bm 1_{k_1k_2}  \right]^2 \\
    \nonumber
    &= \sum_{i=1}^k \sum_{j\in\mathcal{U}} \left[ (GQ)_{ij}(X^T \bm v_Q)_j - \bm 1_{k_1k_2}  \right]^2 \\
    \nonumber
    &= \Vert (GQ)_\mathcal{U} \bm v_Q - \bm 1_{k_1k_2} \Vert_2^2 \\
    \label{eqn:perm_strag-zeroing}
    &=  \Vert (GQ)_\mathcal{U} \bm v_{opt}(GQ, \mathcal{U}) - \bm 1_{k_1k_2} \Vert_2^2.
\end{align}

Moreover, we claim that for any ``zeroing'' matrix Y, there exists $\tilde{\mathcal{U}} \subset [n]$, such that $|\tilde{\mathcal{U}}| = n-s$ and 
\begin{equation}
\label{eqn:zeroing-perm_strag}
    \Vert (G Y) (Y^T \bm v_Q) - \bm 1_{k_1k_2} \Vert_2^2 = \Vert G_{\tilde{\mathcal{U}}} \bm v - \bm 1_{k_1k_2} \Vert_2^2,
\end{equation}
where $\bm v$ is obtained by simply removing all entries of $\bm v_Q$  that are set to zero by $Y$. 

Indeed, define
\begin{equation*}
    \mathcal{\tilde{U}}  = \{j \in [n] : T_{ij}\neq 0 \quad \forall i\in [n]\}
\end{equation*}
In words, $\mathcal{\tilde{U}}$ is the set of workers that are not mapped to the zero vector by $T$. Observe that 
\begin{equation*}
    Y = [\bm e_{\pi(1)} \mathbbm{1}(\pi(1) \in \mathcal{\tilde{U}}), \bm e_{\pi(2)}\mathbbm{1}(\pi(2) \in \mathcal{\tilde{U}}), \dotsc , \bm e_{\pi(n)}\mathbbm{1}(\pi(n) \in \mathcal{\tilde{U}})].
\end{equation*}
Then~\eqref{eqn:zeroing-perm_strag} follows from~\eqref{eqn:perm_strag-zeroing}.

Finally, writing $\tilde{G} := G^{(1)} \otimes G^{(2)}$ and $\hat{G} = G^{(2)} \otimes G^{(1)}$ for convenience, we have
\begin{align*}
    &\phantom{=}\error(G^{(2)} \otimes G^{(1)}) \\
    &= \frac{1}{k_1k_2} \max_{\mathcal{U}} \left\Vert \left( \hat{G} \right)_\mathcal{U} \bm v_\mathrm{opt}(\hat{G}, \mathcal{U}) - \bm 1_{k_1k_2} \right\Vert_2^2 \\
    &= \frac{1}{k_1k_2} \max_{\mathcal{U}} \left\Vert \left( P \left( \tilde{G} \right) Q  \right)_\mathcal{U} \bm v_\mathrm{opt}(\hat{G}, \mathcal{U}) - \bm 1_{k_1k_2} \right\Vert_2^2 \\
    &= \frac{1}{k_1k_2} \max_{\mathcal{U}} \left\Vert \left(\left( \tilde{G} \right) Q  \right)_\mathcal{U} \bm v_\mathrm{opt}(\hat{G}, \mathcal{U}) - \bm 1_{k_1k_2} \right\Vert_2^2 \\
    &\stackrel{(a)}{=} \frac{1}{k_1k_2} \max_{\mathcal{U}} \left\Vert \left( \tilde{G}  \right)_\mathcal{\tilde{U}} \bm v_\mathrm{opt}(\hat{G}, \mathcal{U}) - \bm 1_{k_1k_2} \right\Vert_2^2 \\
    &\le \frac{1}{k_1k_2} \max_{\mathcal{U}} \left\Vert \left( \tilde{G}  \right)_\mathcal{\tilde{U}} \bm v_\mathrm{opt}(\tilde{G}, \mathcal{\tilde{U}}) - \bm 1_{k_1k_2} \right\Vert_2^2 \\
    &\le \frac{1}{k_1k_2} \max_{\mathcal{\tilde{U}}} \left\Vert \left( \tilde{G}  \right)_\mathcal{\tilde{U}} \bm v_\mathrm{opt}(\tilde{G}, \mathcal{\tilde{U}}) - \bm 1_{k_1k_2} \right\Vert_2^2 \\
    &= \error(G^{(1)} \otimes G^{(2)} , s).
\end{align*}
Above, $(a)$ follows from~\eqref{eqn:perm_strag-zeroing} and~\eqref{eqn:zeroing-perm_strag}.

Since there exist permutation matrices $P'$ and $Q'$ such that 
\begin{equation*}
    G^{(1)} \otimes G^{(2)} = P' \left( G^{(2)} \otimes G^{(1)} \right) Q',
\end{equation*}
by the same argument as above, we have
\begin{equation*}
    \error(G^{(1)} \otimes G^{(2)} , s) \le \error(G^{(2)} \otimes G^{(1)} , s).
\end{equation*}

\section{Proof of Lemma~\ref{lemma:err-general-lambda-gc}}
\label{appendix:err-general-lambda-gc}

The proof follows essentially from the proof of Theorem 1 in~\cite{Kadhe--Koyluoglu--Ramchandran2019}, and is included here for completeness.

We first obtain the optimal decoding vector of $G$ for any set of $\tilde{s} := n-s$ non-stragglers $\mathcal{U}$. Recall 
\begin{equation*}
\bm v_\mathrm{opt} (G,\mathcal{U}) := \arg \min_{\bm v \in \mathbbm{R}^{n-s}} \Vert G_\mathcal{U} \bm v - \bm 1_k \Vert_2^2.
\end{equation*}
An optimal solution to the above optimization problem is given by $\bm v_\mathrm{opt} (G,\mathcal{U})  = G_\mathcal{U}^\dagger \bm 1_k$, where $G_\mathcal{U}^\dagger$ is the Moore-Penrose inverse of $G_\mathcal{U}$. It is known that $G_\mathcal{U}^\dagger = (G_\mathcal{U}^T G_\mathcal{U})^{-1} G_\mathcal{U}^T$ when $G_\mathcal{U}^T G_\mathcal{U}$ is invertible. Observe that 
\begin{equation}
\label{eqn:vopt_intermediate_comp}
    G_\mathcal{U}^T G_\mathcal{U} = (l-\lambda) I_{\tilde{s}} + \lambda J_{\tilde{s} \times \tilde{s}}
\end{equation}
since each column of $G_\mathcal{U}$ has $l$ ones, and any pair of columns of $G_\mathcal{U}$ has $\lambda$ intersections. Since $l>\lambda$ by assumption, $G_\mathcal{U}^T G_\mathcal{U}$ is invertible. Thus
\begin{equation*}
    \bm v_\mathrm{opt} (G,\mathcal{U})  = (G_\mathcal{U}^T G_\mathcal{U})^{-1} G_\mathcal{U}^T \bm 1_k. 
\end{equation*}
We have 
\begin{align}
    (G_\mathcal{U}^T G_\mathcal{U})^{-1}
    &= \Big( (l-\lambda) I_{\tilde{s}} + \lambda J_{\tilde{s} \times \tilde{s}} \Big)^{-1} \nonumber \\
    &\stackrel{(a)}{=} \frac{1}{l-\lambda} I_{\tilde{s}} + \frac{1}{1+\trace\left(\frac{\lambda}{l-\lambda} J_{\tilde{s}}\right)} \frac{\lambda}{(l-\lambda)^2} J_{\tilde{s}} \nonumber \\
    \label{eqn:inverse_computation}
    &= \frac{1}{l-\lambda} \left(I_{\tilde{s}} + \frac{\lambda}{l+\lambda(\tilde{s}-1)} J_{\tilde{s}} \right),
\end{align}
where $(a)$ follows from the matrix inversion lemma~\cite{Miller1981}.
Thus the optimal decoding vector is given by 
\begin{align}
    \bm v_\mathrm{opt} (G,\mathcal{U})
    = (G_\mathcal{U}^T G_\mathcal{U})^{-1} G_\mathcal{U}^T \bm 1_k
    \stackrel{(a)}{=} (G_\mathcal{U}^T G_\mathcal{U})^{-1} l \bm 1_{\tilde{s}} 
    \label{eqn:optimal_decoding_vector}
    \stackrel{(b)}{=} \frac{l}{l+\lambda(\tilde{s}-1)} \bm 1_{\tilde{s}},
\end{align}
where $(a)$ follows since each column of $G_\mathcal{U}$ has $l$ ones, and $(b)$ follows from equation~\eqref{eqn:inverse_computation}.

We now find the normalized worst-case squared error of $G$ with $s$ stragglers. Observe that 
\begin{align*}
    &\hspace{1.5em} \Vert G_\mathcal{U} \bm v_\mathrm{opt} (G,\mathcal{U}) - \bm 1_k \Vert_2^2 \\
    &=\left( G_\mathcal{U} \bm v_\mathrm{opt} (G,\mathcal{U}) - \bm 1_k \right)^T \left( G_\mathcal{U} \bm v_\mathrm{opt} (G,\mathcal{U}) - \bm 1_k \right) \\
    &=  \Big(\bm 1_k^T \bm 1_k -2 \bm v_\mathrm{opt}^T (G,\mathcal{U}) G_\mathcal{U}^T \bm 1_k + \bm v_\mathrm{opt}^T (G,\mathcal{U}) G_\mathcal{U}^T G_\mathcal{U} \bm v_\mathrm{opt} (G,\mathcal{U}) \Big)\\
    &\stackrel{(a)}{=} \Big( k - 2l \bm v_\mathrm{opt}^T \bm 1_{\tilde{s}} + \bm v_\mathrm{opt}^T (G,\mathcal{U}) \big( (l-\lambda) I_{\tilde{s}} + \lambda J_{\tilde{s} \times \tilde{s}} \big) \bm v_\mathrm{opt} (G,\mathcal{U}) \Big) \\ 
    &\stackrel{(b)}{=} \left( k - \frac{2 l^2 \tilde{s}}{l+\lambda(\tilde{s}-1)} + \frac{l^2 \left( (l-\lambda)\tilde{s} + (\lambda) \tilde{s}^2 \right)}{(l+\lambda(\tilde{s}-1))^2}  \right) \\
    &= k - \frac{l^2 \tilde{s}}{l+ \lambda (\tilde{s}-1)}.
\end{align*}
where $(a)$ follows since each column of $G_\mathcal{U}$ has $l$ ones, and from equation~\eqref{eqn:vopt_intermediate_comp}. Above, $(b)$ follows from equation~\eqref{eqn:optimal_decoding_vector}. 

\section{Proof of Lemma~\ref{lemma:bibd_extension}}
\label{appendix:bibd_extension}
Notice that $G^\mathrm{B}\otimes \bm 1_{k_2}$ is an $(n_1,k_1k_2,l_1k_2,r_1,\lambda_1k_2)$-GC, thus by Lemma~\ref{lemma:err-general-lambda-gc}
\begin{align*}
    \left\Vert (G^\mathrm{B}\otimes \bm 1_{k_2})_\mathcal{U} v_\mathrm{opt}(G,\mathcal{U}) - \bm 1_{k_1k_2} \right\Vert_2^2 %
    &= k_1k_2 - \frac{(l_1k_2)^2(n_1-s)}{l_1k_2 + \lambda_1k_2(n_1-s-1)} \\
    &= k_1k_2 \left( 1 - \frac{l_1^2(n_1-s)}{l_1k_1 + \lambda_1k_1(n_1-s-1)} \right) \\
    &\stackrel{(a)}{=} k_1k_2 \error(G^B,s),
\end{align*}
where $(a)$ follows from equation~\eqref{eqn:bibd-error}. Observe that $\bm 1_{k_2} \otimes G^\mathrm{B}$ is also an $(n_1,k_1k_2,l_1k_2,r_1,\lambda_1k_2)$-GC, thus 
\begin{align*}
    \left\Vert (\bm 1_{k_2} \otimes G^\mathrm{B})_\mathcal{U} v_\mathrm{opt}(G,\mathcal{U}) - \bm 1_{k_1k_2} \right\Vert_2^2 %
    &= \left\Vert G_\mathcal{U} v_\mathrm{opt}(G,\mathcal{U}) - \bm 1_{k_1k_2} \right\Vert_2^2.
\end{align*}
Dividing by $k_1k_2$ gives the desired equality. 

\section{Proof of Lemma~\ref{lemma:gc_replication}}
\label{appendix:gc_replication}
By construction, the gradient code $\bm 1_{r}^T \otimes G$ is given by
\begin{equation*}
  \begin{bmatrix}
  G^{(1)} & G^{(2)} & \dotsc G^{(r)}
  \end{bmatrix}  
\end{equation*}
where $G^{(i)} = G$ is the $i$-th block of $G$, for each $i\in[r]$. 
We first show that 
\begin{equation*}
    \error(\bm 1_{r}^T\otimes G, s) \leq \error\left( G, \left\lfloor \frac{s}{r} \right\rfloor \right).
\end{equation*}
To this end, consider the decoding vector $\bm v$
that decodes only using the results returned by the block $i^{*}$ with the fewest stragglers, and the best straggling situation among all other blocks with the same number of stragglers. Let block $i\in[r]$ have $s_i$ non stragglers $\mathcal{U}_i$. Then we can write
\begin{equation*}
    \bm v = (\bm v_1^T, \bm v_2^T, \dotsc, \bm v_r^T),
\end{equation*}
where $\bm v_i = \bm 0_{s_i}$ for all $i \neq i^{*}$, and $\bm v_{i^{*}} = v_\mathrm{opt}(G,\mathcal{U}_{i^{*}})$.
Then
\begin{align*}
    \error(\bm 1_{r}^T \otimes G, s) 
    &= \frac{1}{k} \max_{\mathcal{U}} \left\Vert (\bm 1_{r}^T \otimes G)_\mathcal{U} \bm v_\mathrm{opt} - \bm 1_k \right\Vert_2^2 \\
    &\stackrel{(a)}{\leq} \frac{1}{k} \max_{\mathcal{U}} \left\Vert (\bm 1_{r}^T \otimes G)_\mathcal{U} \bm v - \bm 1_k \right\Vert_2^2 \\
    &\stackrel{(b)}{=} \frac{1}{k} \max_{\mathcal{U}} \min_i  \left\Vert G^{(i)}_{\mathcal{U}_i} \bm v_i - \bm 1_k \right\Vert_2^2 \\
    &\stackrel{(c)}{=} \frac{1}{k} \max_{\mathcal{U}} \left\Vert G^{(i^{*})}_{\mathcal{U}_{i^{*}}} \bm v_\mathrm{opt}(G,\mathcal{U}_{i^{*}}) - \bm 1_k \right\Vert_2^2 \\
    &\stackrel{(d)}{\leq} \frac{1}{k} \max_{\mathcal{U}} \left( k \error(G^{(i^{*})},s_{i^{*}}) \right) \\
    &\stackrel{(e)}{=} \error\left(G, \left\lfloor \frac{s}{r} \right\rfloor \right).
\end{align*}
Above, $(a)$ follows since $\bm v$ is not necessarily the optimal decoding vector. Secondly, $(b)$ follows from the choice of decoding vector $\bm v$, and $(c)$ follows since $\bm v_i = v_\mathrm{opt}(G,\mathcal{U}_i)$ for the block with the least error. Moreover, $(d)$ follows since $\mathcal{U}_{i^{*}}$ may not be the worst-case set of non stragglers for block $i^{*}$. Lastly, $(e)$ follows since the error of $G$ is monotone increasing, and by observing that the error of the lowest error block $i^{*}$ is maximized when each block has at least $\lfloor s/r \rfloor$ stragglers.

We now show that 
\begin{equation*}
    \error(\bm 1_{r}^T\otimes G, s) \ge \error\left( G, \left\lfloor \frac{s}{r} \right\rfloor \right).
\end{equation*}
To this end, for each $z\in[n]$, let $\mathcal{A}_z$ be a worst-case set of $n-z$ non-stragglers of $G$. Now, let $\mathcal{B}$ be the set of $nr-s$ non-stragglers of $\bm 1_r^T \otimes G$ given by 
\begin{equation*}
    \mathcal{B} = \bigcup_{i=0}^{r-1} \left\{\mathcal{A}_{\lfloor s/r \rfloor} + ni \right\} ,
\end{equation*}
where for any $z$, $\mathcal{A}_z + ni = \{ a + ni: a \in \mathcal{A}_z \}$.
We can interpret $\mathcal{B}$ as the set of non-stragglers when each copy of $G$ has $\lfloor \tfrac{s}{r} \rfloor$ stragglers in the worst case pattern, and the remaining $s - \lfloor \tfrac{s}{r} \rfloor r$ stragglers are placed arbitrarily. Note that there will be at one block with $\lfloor \tfrac{s}{r} \rfloor r$ stragglers in the worst-case straggling pattern. Then
\begin{align*}
    \error(\bm 1_{r}^T \otimes G, s) 
    &= \frac{1}{k} \max_{\mathcal{U}} \left\Vert (\bm 1_{r}^T \otimes G)_\mathcal{U} \bm v_\mathrm{opt} - \bm 1_k \right\Vert_2^2 \\
    &\stackrel{(a)}{\geq} \frac{1}{k} \left\Vert (\bm 1_{r}^T \otimes G)_\mathcal{B} \bm v_\mathrm{opt} - \bm 1_k \right\Vert_2^2 \\
    &\stackrel{(b)}{=} \frac{1}{k} \left\Vert G_{\mathcal{A}_{\lfloor s/r \rfloor}} \bm v_\mathrm{opt}(G,\mathcal{A}_{\lfloor s/r \rfloor}) - \bm 1_k \right\Vert_2^2 \\
    &\stackrel{(c)}{=} \error\left( G, \left\lfloor \frac{s}{r} \right\rfloor \right).
\end{align*}
Above, $(a)$ follows since $\mathcal{B}$ may not be the worst-case set of non stragglers. Moreover, $(b)$ follows by observing that if $\bm 1_r^T \otimes G$ has non-stragglers $\mathcal{B}$, then by construction each identical block has the same set of non-stragglers $\mathcal{A}_{\lfloor s/r \rfloor}$. Therefore, optimally decoding $(\bm 1_r^T \otimes G)_\mathcal{B}$ is the same as optimally decoding any block. Lastly, $(c)$ follows since $\mathcal{A}_{\lfloor s/r \rfloor}$ is the worst-case set of $\lfloor s/r \rfloor$ non-stragglers of $G$.

\section{Proof of Lemma~\ref{lemma:block_diag_norm}}
\label{appendix:block_diag_norm}
The workers and gradients of $G^{(j)}$ and $G^{(j')}$ are disjoint if $j\ne j'$ for any $1\le j,j'\le n\tau$. Then 
\begin{equation*}
    \Vert G_\mathcal{U} \bm v - \bm 1_{k\tau} \Vert_2^2 
    = \sum_{i=1}^{k\tau} \left((G_\mathcal{U} \bm v)_i - 1 \right)^2 
    = \sum_{j=1}^{\tau} \sum_{i=1}^{k} \left((G^{(j)}_{\mathcal{U}_j} \bm v_j)_i - 1 \right)^2 = \sum_{j=1}^\tau \Vert G^{(j)}_{\mathcal{U}_j} \bm v_j - \bm 1_{k} \Vert_2^2.
\end{equation*}

\section{Proof of Lemma~\ref{lemma:convexity_bibd_error}}
\label{appendix:convexity_bibd_error}
We first show that if a function $f:\mathbbm{R}\to\mathbbm{R}$ is convex, then $\{f(m)\}_{m=0}^\infty$ is a convex sequence. If $f$ is convex, then for all $t\in[0,1]$ and any $x_1,x_2\in\mathbbm{R}$
\[
f( tx_1 + (1-t)x_2) \le tf(x_1) + (1-t)f(x_2) .
\]
Set $t=\frac{1}{2}$, and let $x_1 = m-1$ and $x_2=m+1$ for any $m\in\mathbbm{N}$. Then the above expression becomes
\[
f(m) \leq \frac{f(m-1) + f(m+1)}{2},
\]
thus $\{f(m)\}_{m=0}^\infty$ is a convex sequence. It remains to show that the BIBD error expression is a convex function. Let $f:[0,n]\to\mathbbm{R}$ be the function given by 
\[
f(x) = 1 - \frac{l^2(n-x)}{kl+k\lambda(n-x-1)}
\]
for any $x\in[0,n]$, and notice that $\{\error(G^\mathrm{B},s)\}_{s=0}^n = \{f(s)\}_{s=0}^n$. Since 
\[
\frac{\partial^2 f}{\partial x ^2 } = \frac{2l^2\lambda(\lambda-l)}{k(\lambda(x-n-1)+1)}
\]
is non-negative for any $x\in[0,n]$, $f$ is a convex function, and $\{\error(G^\mathrm{B},s)\}_{s=0}^n$ is a convex sequence.

\section{Proof of Lemma~\ref{lemma:err_kron_bibd}}
\label{appendix:err_kron_bibd}
We write 
\begin{equation*}
    G_\mathcal{U} = 
    \begin{bmatrix}
    \vert & \vert & \dotsc & \vert \\
    \bm w_1 & \bm w_2 & \dotsc &  \bm w_{\tilde{s}} \\
    \vert & \vert & \dotsc & \vert 
\end{bmatrix},
\end{equation*}
where $\bm w_1, \bm w_2,\dotsc, \bm w_{\tilde{s}}$ are column vectors of size $k_1k_2$. Then 
$(G_\mathcal{U}^T  G_\mathcal{U})_{i,j} = \bm w_i^T \bm w_j$
is the number of intersections between non-straggling workers $i,j\in\mathcal{U}$. To determine the number of intersections between workers $i$ and $j$, we partition the $[n_1n_2]$ workers of $G$ into ``blocks'' and ``classes''. 
Let block $m\in[n_1]$ refer to the set of $n_2$ workers 
\begin{equation*}
    \mathcal{B}_m := \{(m-1)n_2+1,(m-1)n_2+2,\dotsc,m n_2\}. 
\end{equation*}
Similarly, class $m'\in[n_1]$ refers to the set of $n_1$ workers
\begin{equation*}
    \mathcal{C}_{m'} := \{m', m'+n_2, m'+2n_2,\dotsc, m'+{n_1-1}n_2\}.
\end{equation*}
Let $i_\mathrm{b}$ and $j_\mathrm{b}$ be the block  that $i$ and $j$ are from respectively i.e. $i\in \mathcal{B}_{i_\mathrm{b}}$ and $j\in \mathcal{B}_{j_\mathrm{b}}$. Similarly, we denote by $i_\mathrm{c}$ and $j_\mathrm{c}$ the classes that workers $i$ and $j$ are in. By construction of $G$, notice that
\begin{equation*}
    \bm w_i^T \bm w_j = 
    \begin{cases} 
      l_1l_2 & i=j, \\
      l_1\lambda_2 & i\neq j, i_\mathrm{b} = j_\mathrm{b}, \\
      \lambda_1l_2 & i\neq j, i_\mathrm{c} = j_\mathrm{c}, \\ 
      \lambda_1\lambda_2 & \text{otherwise.}
   \end{cases}
\end{equation*}
Indeed, this follows by observing that if $i=j$, then $i_b = j_b$, and $i_c = j_c$, thus the number of intersections between workers $i$ and $j$ is $l_1l_2$. Similarly, if $i_b = j_b$ and $i \neq j$, then workers $i$ and $j$ correspond to the same column in $G^\mathrm{PB1}$ but different columns in $G^\mathrm{PB2}$. Moreover, if $i_b = j_b$ and $i \neq j$, then workers $i$ and $j$ correspond to the same column in $G^\mathrm{PB2}$ but different columns in $G^\mathrm{PB1}$. Lastly, if workers $i\neq j$ are neither from the same block nor the same class, then $i$ and $j$ correspond to different columns of both $G^\mathrm{PB1}$ and $G^\mathrm{PB2}$.

Let $\tilde{s}_{i_\mathrm{b}} = \left|\mathcal{U}\cap \mathcal{B}_{i_\mathrm{b}} \right|$ be the number of non-straggling workers in block $i_\mathrm{b}$, and $\tilde{z}_{i_\mathrm{c}} = \left |\mathcal{U}\cap \mathcal{C}_{i_\mathrm{c}} \right|$ be the number of non-straggling workers in class $i_\mathrm{c}$. Then 
\begin{align*}
    &\hspace{1.5em} 1_{\tilde{s}}^T G_\mathcal{U}^T  G_\mathcal{U} \bm 1_{\tilde{s}} \\
    &= \sum_{i\in\mathcal{U}} \sum_{j\in\mathcal{U}} \bm w_i^T \bm w_j \\
    &= \sum_{i\in\mathcal{U}} \Big( l_1l_2 + (\tilde{s}_{i_\mathrm{b}}-1)l_1\lambda_2 + (\tilde{z}_{i_\mathrm{c}}-1)(\lambda_1l_2) + (\tilde{s} - \tilde{s}_{i_\mathrm{b}} - \tilde{z}_{i_\mathrm{c}} + 1)\lambda_1\lambda_2 \Big)\\
    &= (l_1-\lambda_1)(l_2-\lambda_2)\tilde{s} + \lambda_1\lambda_2\tilde{s}^2 + (l_1\lambda_2 - \lambda_1\lambda_2)\sum_{i\in\mathcal{U}} \tilde{s}_{i_\mathrm{b}}
    + (l_2\lambda_1 - \lambda_1\lambda_2)\sum_{i\in\mathcal{U}} \tilde{z}_{i_\mathrm{c}}. 
\end{align*}
Now observe that 
\begin{equation*}
    \sum_{i\in\mathcal{U}} \tilde{s}_{i_\mathrm{b}} = \sum_{m=1}^{n_1}\sum_{i\in\mathcal{U}\cap\mathcal{B}_m} \tilde{s}_{i_\mathrm{b}} 
    = \sum_{m=1}^{n_1}\sum_{i\in\mathcal{U}\cap\mathcal{B}_m} \tilde{s}_{m} 
    = \sum_{m=1}^{n_1} \tilde{s}_{m}^2. 
\end{equation*}
Notice that $0\leq\tilde{s}_{m}\leq n_2$ for each $m\in[n_1]$, and $\sum_{m=1}^{n_1} \tilde{s}_{m} = \tilde{s}$. Therefore
\begin{equation*}
    \sum_{i\in\mathcal{U}} \tilde{s}_{i_\mathrm{b}} = \sum_{m=1}^{n_1} \tilde{s}_{m}^2 \leq \sum_{m=1}^{n_1} \tilde{s}_{m} n_2 = n_2 \tilde{s}. 
\end{equation*}
Similarly, we have 
\begin{equation*}
    \sum_{i\in\mathcal{U}} \tilde{z}_{i_\mathrm{c}} = \sum_{m'=1}^{n_2} \sum_{i\in\mathcal{U}\cap\mathcal{C}_{m'}} \tilde{z}_{i_\mathrm{c}} 
    = \sum_{m'=1}^{n_2} \sum_{i\in\mathcal{U}\cap\mathcal{C}_{m'}} \tilde{z}_{m'} 
    = \sum_{m'=1}^{n_2} \tilde{z}_{m'}^2 
    \stackrel{(a)}{\le} \sum_{m'=1}^{n_2} \tilde{z}_{m'} n_1 
    \stackrel{(b)}{=} n_1 \tilde{s}.
\end{equation*}
where $(a)$ follows since $0 \le \tilde{z}_{m'} \le n_1$ for any $m'\in[n_2]$, and $(b)$ follows since $\tilde{z}_{1} + \tilde{z}_{2}+\dotsc + \tilde{z}_{n_2} = \tilde{s}$. 
Therefore, we have 
\begin{equation*}
    \bm 1_{\tilde{s}}^T G_\mathcal{U}^T  G_\mathcal{U} \bm 1_{\tilde{s}} \le \tilde{s}d + \lambda_1\lambda_2 \tilde{s}^2. 
\end{equation*}

\section*{Acknowledgment}
The authors would like to thank Ziqiao Lin, who participated as a summer intern student in the initial stage of this work. 
This work was supported in part by the NSERC Discovery Launch Supplement DGECR-2019-00447, in part by the NSERC Discovery Grant RGPIN-2019-05448, and in part by the NSERC Collaborative Research and Development Grant CRDPJ 543676-19.

\end{document}